\documentclass[ida, crcready]{iosart2x}

\usepackage{siunitx}

\begin{document}

\begin{frontmatter}

%\pretitle{}
\title{A Time Series Approach To Player Churn and Conversion in Videogames}
\runningtitle{A Time Series Approach To Player Churn and Conversion in Videogames}

\author[A,B]{\inits{A.}\fnms{Ana} \snm{Fern\'andez del R\'io}\ead[label=e1]{ana.fernandezdelrio@invi.uned.es}%
\thanks{Corresponding author. \printead{e1}.}},
\author[A]{\inits{A.}\fnms{Anna} \snm{Guitart}\ead[label=e2]{}}
and
\author[A]{\inits{A.}\fnms{\'Africa} \snm{Peri\'an\~ez}\ead[label=e3]{}}
\runningauthor{A. Fern\'andez del R\'io et al.}
\address[A]{Data Science Research, \orgname{Yokozuna Data, a Keywords Studio},Tokyo, \cny{Japan}}
\address[B]{Departamento de F\'isica Fundamental, \orgname{Universidad Nacional de Educaci\'on a Distancia (UNED)}, Madrid, \cny{Spain}}

\begin{abstract}
 Players of a free-to-play game are divided into three main groups: non-paying active users, paying active users and inactive users. A State Space time series approach is then used to model the daily conversion rates between the different groups, i.e., the probability of transitioning from one group to another. This allows, not only for predictions on how these rates are to evolve, but also for a deeper understanding of the impact that in-game planning and calendar effects have. It is also used in this work for the detection of marketing and promotion campaigns about which no information is available. In particular, two different State Space formulations are considered and compared: an Autoregressive Integrated Moving Average process and an Unobserved Components approach, in both cases with a linear regression to explanatory variables. Both yield very close estimations for covariate parameters, producing forecasts with similar performances for most transition rates. While the Unobserved Components approach is more robust and needs less human intervention in regards to model definition, it produces significantly worse forecasts for non-paying user abandonment probability. More critically, it also fails to detect a plausible marketing and promotion campaign scenario.

\end{abstract}


\end{frontmatter}

%%%%%%%%%%%%%%%%%%%%%%%%%%%%%%%%%%%%%%%%%%%%%%%%%%%%%%%%%%%%%%%%%%%%%%%%%%%%%%%%%%%%%%%%%%
\section{Introduction}
\label{sec:intro}

Player profiling has become of crucial importance for the video game industry in an increasingly competitive market. For many titles, and specially for free-to-play-games, the main source of revenue are in-app purchases \cite{Monetization}. Thus, characterizing players based on their purchasing behavior is a common practice when aiming to improve game monetization. In this study, the total player population is divided into three groups: paying users (PUs), non-paying users (non-PUs) and inactive players, i.e., those that have already abandoned the game or \emph{churned}. This classification is not as straight forward as it could appear for most titles. The definition of churn and of who is or is not a PU, are in themselves, as for any service not bounded by contract, not clear for most video games, as is discussed for example in Guitart et al. (2019) \cite{guitart2019understanding}.

 The goal of this work is to understand and predict the evolution of the daily conversion or transition rates from one group to another. In particular, the focus is on churn probability (for both PUs and non-PUs), on non-PU to PU transition rate (sometimes refered to in the literarture simply as \emph{player conversion}), and on the the probability of PU becoming non-PU (or \emph{purchase churn}). These will be modeled using time series State Space Models (SSMs), taking as covariates or explanatory regressors information regarding in-game planning, as well as holidays and other calendar effects. The main goal of this work is then to study the evolution of probability for a player of any group to transition to another one (and thus also
of remaining in the same group). These models can be used to predict the daily conversion rates between groups. They also provide a measure of the impact of the different covariates, thus yielding a classification of, for example, in-game events, depending on which transition probabilities they impact, with which sign, and how large the effect is. 

There would be several practical uses of having such a system running operationally. Accurate forecasts could be used to improve resource allocation. While these could be useful, predictions are in themselves probably not the most interesting application in this case. Having however an estimation of the impact of different events (inside the game and external), could significantly improve in-game planning. Finally, as it will be described, this type of modelling can also be used to try to detect missing information that would correspond to large discrepancies between model and reality. In this work, this approach is aimed mainly at the detection of new user acquisition campaigns and promotion campaigns (for which there is no information available except that they are known to have existed often and with significant impact). In a production setup, where all relevant information would be available, it could be used as some sort of automatic monitoring, to detect for example server failures or buggy new releases. It would also allow for comparison of how an event going on with all its particularities compares (positively or negatively) to the average effect that type of event has had in the past, i.e., to measure the relative success of any individual event or campaign.   

Another interesting path to explore would be to use these predictions in the modeling of individual player behavior, for example for churn predictions \cite{perianez2016churn,GameBigData,cig2018competition,chen2019competition,guitart2019understanding} or of conversion to PU \cite{guitart2019non}. By using them as features, the effect of calendar effects, campaigns and in-game events could be easily incorporated, and the models thought of as \emph{correcting} the probability of a phenomenon (churn, conversion\ldots) in each group to reflect the probability of each individual. 

Of course, this approach can be extended to more complex landscapes of user types. Players can be further divided into  additional subgroups depending on their specific purchasing behavior (distinguishing for example, between frequent and impulsive purchasers, and/or between top spenders and the rest), skills, play frequency and/or playstyle; and transitions between all these groups studied with this technique. This could provide deep insights about game dynamics and how different types of players are affected differently by different game planning strategies and events outside the game.

While the setup under study is very specific, this approach could be useful outside the realm of videogames. It is directly translatable with identical user grouping to any online platform where purchases are available. It could also be applied to online communities to understand, for example, how users that produce content are affected differently by events than those that merely consume it, and what drives the transition between these two regimes. Even in physical stores a similar approach could be used to model the probability that someone coming into the store will make a purchase. It is basically a valid approach in any setting where there are users or customers of different types, and where it can be interesting to analyze and/or predict how users transition from one type to another.  

 To the best of our knowledge, this is the first work in video-games that predicts transition probabilities between different types of players and provides an analysis of the impact in-game and calendar events have in them.
 
 This paper is organized as follows: Section \ref{sec:related} compiles some relevant bibliographic references of both related statistical and machine learning applications to videogames and of time series approaches in other fields. The models used are described in Section \ref{sec:ssm}, and the particular dataset to which these are applied to in Section \ref{sec:data}. The methodology followed is described in detail in Section \ref{sec:meth}, and the results are then presented and discussed in Section \ref{sec:results}. The paper closes with a summary and conclusions in Section \ref{sec:conc} and a brief description of the software used in \ref{sec:software}.

%%%%%%%%%%%%%%%%%%%%%%%%%%%%%%%%%%%%%%%%%%%%%%%%%%%%%%%%%%%%%%%%%%%%%%%%%%%%%%%%%%%%%%%%%%
\section{Related works}
\label{sec:related}

 There is not much work available in terms of time-series prediction in video-games. In Guitart et al. (2017) \cite{guitart2017forecasting} the total aggregated value of daily sales and playtime is forecasted via different time series modelling techniques. Online traffic generated by on-line first person shooter games is dealt with from a time series perspective in Cricenti et al. \cite{cricenti2007time}. Previous attempts at player profiling include for example Drachen et al. \cite{drachen2012guns}, Saas et al. \cite{saas2016discovering}, Fern\'andez del R\'io et al. \cite{del2019profiling} and Guitart et al. (2019) \cite{guitart2019understanding}. Numerous previous studies on in-game player behavior have focused on individual player predictions (and not on aggregated time series as is the case of the present work). For example, in determining which players are going to abandon the game and when \cite{perianez2016churn,GameBigData,cig2018competition,chen2019competition,guitart2019understanding}, predicting their lifetime-value \cite{sifa2018customer,drachen2018or,chen2018ltv} and their purchase decisions \cite{sifa2015predicting, itemPrediction2018}, or if and when players will become PUs \cite{guitart2019non}.

Time series state space modeling approaches are ubiquitous in the study of economic and social processes, both in academia and in the industry. Examples include crime \cite{chen2008}, printed newspaper \cite{permatasari2018} or automobile \cite{shakti2017} demand, stock prices \cite{adebiyi2014, mondal2014}, electricity prices \cite{contreras2003} or epidemics \cite{zhang2014, ho2015, zeng2016}. Applications go beyond human related processes, and they have also been used to predict for example population growth of animal species \cite{bjornstad2001, tolimieru2017} or their parasites \cite{elghafghug}, weather \cite{tektas2010} or water quality \cite{faruk2010}.

%%%%%%%%%%%%%%%%%%%%%%%%%%%%%%%%%%%%%%%%%%%%%%%%%%%%%%%%%%%%%%%%%%%%%%%%%%%%%%%%%%%%%%%%%%
\section{Time series state space modeling}
\label{sec:ssm}

\emph{State Space Models (SSM)} are a broad type of time series models that assume probabilistic dependence between the latent state variable and the observed measurement, thus estimating the state of an unobservable process (latent state) from an observed data set \cite{boxjen76,brockwell, hamilton, shumway2010}. An earlier example of a SSM is the \emph{Kalman Filter (KM)} \cite{kalman1960new}, widely used today. 

 SSMs are made up of two components. The \emph{state transition equation} describes the evolution of the so called latent state $l_{t} \in {\rm I\!R}^{L}$. The \emph{observation model} describes the relation between the latent (non directly observable) state and the time series of observations $z_{t} \in {\rm I\!R}$. The state transition equation describes stochastic transition dynamics and the observation model is also probabilistic in nature, so any such model would be determined by the equations $p(l_{t}|l_{t-1})$ and $p(z_{t}|l_{t})$.

Any linear SSM can be expressed in the form:

\begin{eqnarray}
\label{eq:ssm} 
 l_{t} & = & T_{t}l_{t-1} + c_{t} + R_{t}\eta_{t} \\
 z_{t} & = & D_{t}l_{t} + d_{t} + \epsilon_{t}  
\end{eqnarray}

\noindent where $T_{t}$ is the \emph{transition matrix}, $c_{t}$ the \emph{latent state intercept}, $R_{t}$ the \emph{selection matrix}, $D_{t}$ the \emph{design matrix} and $d_{t}$ the \emph{observation intercept}. The terms $\eta_{t}$ and $\epsilon_{t}$ represent random innovations that are typically considered to be normally distributed, i.e, 

\begin{eqnarray}
 \eta_{t}  & = & \mathcal{N}(0, \Sigma_{t}^{s})\\
    \epsilon_{t} & = & \mathcal{N}(0, \Sigma_{t}^{o})  
\end{eqnarray}

\noindent where $\Sigma_{t}^{s}$ is the \emph{state covariance matrix} and  $\Sigma_{t}^{o}$ is the \emph{observation covariance matrix}.

Many different well known time series models can be described as SSMs. In particular, the two approaches we compare in this paper: \emph{Autoregressive Integrated Moving Average (ARIMA)} and \emph{Unobserved Components (UC)} models (both with regressors) have an SSM formulation \cite{boxjen76,brockwell, hamilton, shumway2010}. Besides, any two given SSMs can be combined. For example:

\begin{eqnarray}
\label{eq:2ssm}
         z_{t} & = & D_{1}l_{1, t} + D_{2}l_{2, t} \\
         l_{1,t} & = & T_{1}l_{1, t-1} + \eta_{1,t}\\
         l_{2,t} & = & T_{2}l_{2, t-1} + \eta_{2, t}
\end{eqnarray}

\noindent would become:

\begin{eqnarray}
\label{eq:2ssm2}
         z_{t} & = &( D_{1} D_{2} ) 
         \begin{pmatrix}
            l_{1,t}\\
            l_{2, t} 
        \end{pmatrix}
        +   \epsilon_{t}\\
      \begin{pmatrix}
            l_{1,t}\\
            l_{2, t} 
      \end{pmatrix} & = &
      \begin{pmatrix}
            T_{1} & 0\\
            0 & T_{2} 
      \end{pmatrix}
      \begin{pmatrix}
            z_{1,t-1}\\
            z_{2,t-2} 
      \end{pmatrix}
              +   \begin{pmatrix}
            \eta_{1,t}\\
            \eta_{2, t} 
        \end{pmatrix}
 \end{eqnarray}

This allows for the combination of a linear regression in the explanatory variables (described as SSM in section \ref{sec:linreg}) with either an ARIMA (described in section \ref{sec:arima}) or an UC (described in section \ref{sec:uc}) stochastic component. While a huge variety of filters and smoothers that can be written using state linear gaussian space model formulation, this section will only describe in some detail the three aforementioned big model families that will be used throughout this paper.

%%%%%%%%%%%%%%%%%%%%%%%%%%%%%%%%%%%%%%%%%%%%%%%%%%%%%%%%%%%%%%%%%%%%%%%%%%%%%%%%%%%%%%%%%%
\subsection{Linear regression}
\label{sec:linreg}

One of the main goals of this work is to understand the deterministic behavior of the series that can be modelled in terms of other exogenous explanatory variables or covariates. This will be done through a linear regression, that can also be expressed as an SSM by setting $T_{t}=d_{t}=R_{t} =0$ and $c_{t}=\sum_{i}\beta_{i}x_{t}^{i}$ in equations \ref{eq:ssm}, where $x^{i}$ are the covariates or regressors and $\beta_{i}$ the corresponding parameters to be estimated.

%%%%%%%%%%%%%%%%%%%%%%%%%%%%%%%%%%%%%%%%%%%%%%%%%%%%%%%%%%%%%%%%%%%%%%%%%%%%%%%%%%%%%%%%%%
\subsection{Autoregressive Integrated Moving Average models}
\label{sec:arima}

An ARIMA model of order (p, d, q) is and stochastic time series model. Each observation has a weighted dependence on the previous $p$ observations (AR terms) and on the previous $q$ noise realizations (MA terms). This results in a parsimonious model in which dependence of each time step to virtually infinite previous lags can be captured with a few parameters. The \emph{integrated} refers to the $d$ differences that can be taken on the original series in order to make it stationary and/or reduce its variance.

An ARMA model (identical to the ARIMA described above without differentiating the time series) with regressors in its better known form is usually written \cite{boxjen76}:

\begin{eqnarray}
\label{eq:arimax}
     z_{t} & = &\alpha + \sum_{i=1}^{n} \beta_{i}x_{i, t} + y_{t}   \\
     y_{t} & =  &\phi_{1}y_{t-1} + \phi_{2}y_{t-2} + \ldots + \phi_{p}y_{t-p} + \\ & & \theta_{1}\epsilon_{t-1} + \theta_{2}\epsilon_{t-2} + \ldots + \theta_{q}\epsilon_{t-q} + \epsilon_{t} \\
     \epsilon_{t} & = & \mathcal{N}(0, \sigma^{2})  
\end{eqnarray}

\noindent where $p$ and $q$ are the orders of the autoregressive (AR) and moving average  (MA) polynomials respectively, $\phi_{1}, \ldots \phi_{p}$ the autorregresive parameters, $\theta_{1}, \ldots \theta_{q}$ the moving average parameters, the $n$ $x_{i}$ are the explanatory variables (covariates or regresors) and the $\beta_{i}$ their associated parameters and $\alpha$ the model's intercept. 

In SSM format, the ARMA(p,q) equation can be written as \cite{brockwell}:

\begin{eqnarray}
        y_{t} & = & (1, 0,\ldots,0)l_{t}\\
        l_{t}  & = &
        \begin{pmatrix}
        \phi_{1} & 1 & 0 & \ldots & 0\\
        \phi_{2} & 0 & 1 & 0 & \ldots \\
        \vdots & \vdots & \vdots & \vdots & \\
        \phi_{r} & 0 & 0 & \ldots & 0
        \end{pmatrix}    l_{t-1}    + 
        \begin{pmatrix}
            1\\
            \theta_{1} \\
            \vdots \\
            \theta_{r}
        \end{pmatrix}
\end{eqnarray}

\noindent where $r=max(p,q+1)$, $\theta_{i}=0$ for $q<i\leq r$ and  $\phi_{i}=0$ for $p<i\leq r$, and $l_{t}^{T}=(y_{t}, y_{t-1}, \ldots y_{t-p})$.

%%%%%%%%%%%%%%%%%%%%%%%%%%%%%%%%%%%%%%%%%%%%%%%%%%%%%%%%%%%%%%%%%%%%%%%%%%%%%%%%%%%%%%%%%%
\subsection{Unobserved Components models}
\label{sec:uc}

We refer to \emph{Unobserved Component} or \emph{Structural Time Series} models to formulations in which a time series is explained in terms of underlying trends, cycles or seasonal dependencies. They can be generally expressed as \cite{harvey1991, harvey2006}:

\begin{equation}
    z_{t} = \mu_{t} + \gamma_{t} + c_{t} + \epsilon_{t}
\end{equation}

\noindent where $\mu_{t}$ is the \emph{trend} component, $\gamma_{t}$ is the \emph{seasonal} component, $c_{t}$ the \emph{cyclic} component and $\epsilon_{t}$ a random shock $\epsilon_{t} \sim \mathcal{N}(0, \sigma^{2})$.

Both the cyclical and seasonal components intend to capture behavior that repeats itself. The seasonal part with a fixed, defined frequency $s$ (for example $s=7$ for weekly seasonality of a daily time series):

\begin{equation}
 \gamma_{t} = - \sum_{j=1}^{s-1}\gamma_{t-j}+w_{t}
 \end{equation}
 
\noindent where $w_{t}$ is random noise with zero mean and variance estimated as an additional parameter. The cyclical through longer periods of unknown frequency:

\begin{eqnarray}
       c_{t+1} & = & c_{t}cos\lambda_{c} + c_{t}^{*}sin\lambda_{c}+u_{t} \\
       c_{t+1}^{*} & = & -c_{t}sin\lambda_{c} + c_{t}^{*}cos\lambda_{c}+u_{t}^{*}
 \end{eqnarray}

\noindent where $u_{t}$ is also normally distributed with mean zero and estimated variance. The cyclic frequency $\lambda$ is also estimated as a parameter.

The trend component can be expressed as:

\begin{eqnarray}
\label{eq:level}
         \mu_{t+1} & = & \mu_{t} +\nu_{t} + \eta_{t+1}\\
         \nu_{t+1} & = & \nu_{t} + \zeta_{t+1}   \label{eq:slope}    
\end{eqnarray}

\noindent where $\eta_{t}$ and $\zeta_{t}$ represent white noise (normally distributed with zero mean) with variances additional parameters to be estimated. If all the elements in Eqs. (\ref{eq:level}) and (\ref{eq:slope}) are non zero the term is referred to as local linear (stochastic) trend. Other particular behaviors correspond to some of the elements of the equation being null: smooth trend ($\eta_{t}=0$), local (stochastic) level and deterministic trend ($\zeta_{t}=0$), deterministic trend ($\eta_{t}=\zeta_{t}=0$), local (stochastic) level ($\nu_{t}=\zeta_{t}=0$) or a simple constant term ($\nu_{t}=\zeta_{t}=\eta_{t}=0$). As will be soon discussed, the local level model is used extensively in this paper. Note that this corresponds to a random walk. 

Many structural time series models are related to ARIMA ones. As the UC formulation typically has several random noise terms, these have to be combined to be made equivalent to the single noise term of ARIMA formulations. This normally translates into some regions of the ARIMA parameter space being forbidden, with this resulting form, equivalent to the UC model, usually refered to as \emph{reduced model} \cite{harvey2006}. For example, the reduced model of a local level is an ARIMA of order (0, 1, 1).

%%%%%%%%%%%%%%%%%%%%%%%%%%%%%%%%%%%%%%%%%%%%%%%%%%%%%%%%%%%%%%%%%%%%%%%%%%%%%%%%%%%%%%%%%%
\section{Dataset}
\label{sec:data}

The game under study is \emph{Age of Ishtaria}, a mobile role-playing card freemium game developed by Silicon Studio. Data is available since its launch on September 25, 2014 to May 9, 2017. In these close to first two years and a half of its history 2107166 players went through the game, of which 33194 did at least one purchase. The game had in this period typically between ten and twenty thousand daily active users (DAU), with peaks (presumably due mainly to new user acquisition campaigns) of nearly fifty thousand DAU.  At the end of the period 18483 players were considered to still be active.

Three groups of players will be the main focus of attention in this work: \emph{non-paying users} (active and not purchasing), \emph{paying users} (active and purchasing) and \emph{churned players} (inactive players).

As it has been mentioned already in section \ref{sec:intro} the definition of churn is not straight forward for online games. Following the method discussed in Guitart et al. (2019) \cite{guitart2019understanding}, players can be considered inactive when they have not logged in for a fixed number of days. This number is determined so as to be useful in detecting churn as soon as possible, while keeping \emph{false churners} (players flagged as churned that come back to the game) and \emph{missed sales} (purchases made by false churners after they come back to the game) under a reasonable threshold. In particular, for this dataset, using the first two months of data, the churn definition is set to 9 days, as this yields less than 10\% false churners
and less than 1.5\% lost sales. Unlike in some previous work related to player purchasing behavior, where all players that have made at least one purchase in their lifetime are considered as PUs, here we also consider transitions between active PU and active non-PU. That is, we consider paying players become non-PUs after a long enough period with no purchasing activity. \emph{Purchase churn} is defined analogously to login churn, and the period without purchases to mark a previous PU as transitioned back to non-PU is set in this case to 50 days.

Data collected includes individual player-related information such as daily logs into the game and purchases made, and also non-user related such as in-game events. The latter are included in the modelling as exogenous variables, considering all in-game event types provided: Gigant Break, Gift Event, Gacha, Duel Arena, Battle Arena, Battle Event, Mission Event, Mission Bingo, Raid Event, Raid Boss, Item Collection, Poll Event, Call to Arms, Raid Battle and Adveniment. The start date and end date of each event of each of these types throughout the period is known, as well as a measure of the expected impact they had when planned, tagged as 0, 1, 2, 3 or 4. Although numeric, it is better understood as a qualitative measure of expected outcome. 

When discussing online videogame data, it is important to note the exceptional quality these datasets have. Every action every player takes in the game is recorded. This makes both the information on daily logs and purchases virtually noise free and eliminates the problem of missing value treatment. Regarding the logs, even if some of the actions were not recorded due to some technical problem, a single action per player would be enough to rightfully count them as logged in that day. In what concerns purchases, players would complain if these were not effective (and thus not recorded). This makes the amount of lost logins and purchases in the dataset negligible. 

Aggregating log and purchase information the following daily time series of populations of interest can be obtained: \emph{PUs} (number of players that have made a purchase in the last 50 days and have logged in in the previous 9 days), \emph{non-PUs}: (number of players that have logged in the past 9 days but have made no purchase in the last 50 days) and {inactive players} (players that have not logged into the game in the previous 9 days).

\begin{figure}
\centering
\includegraphics[width=4cm]{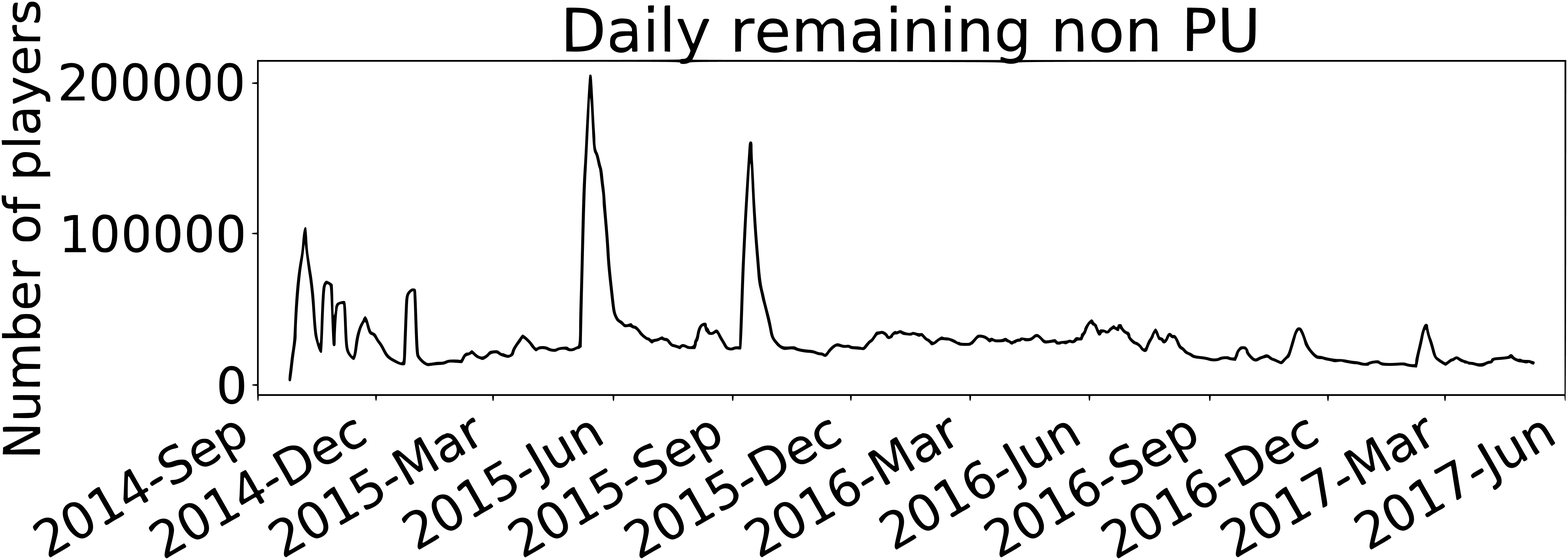}
\includegraphics[width=4cm]{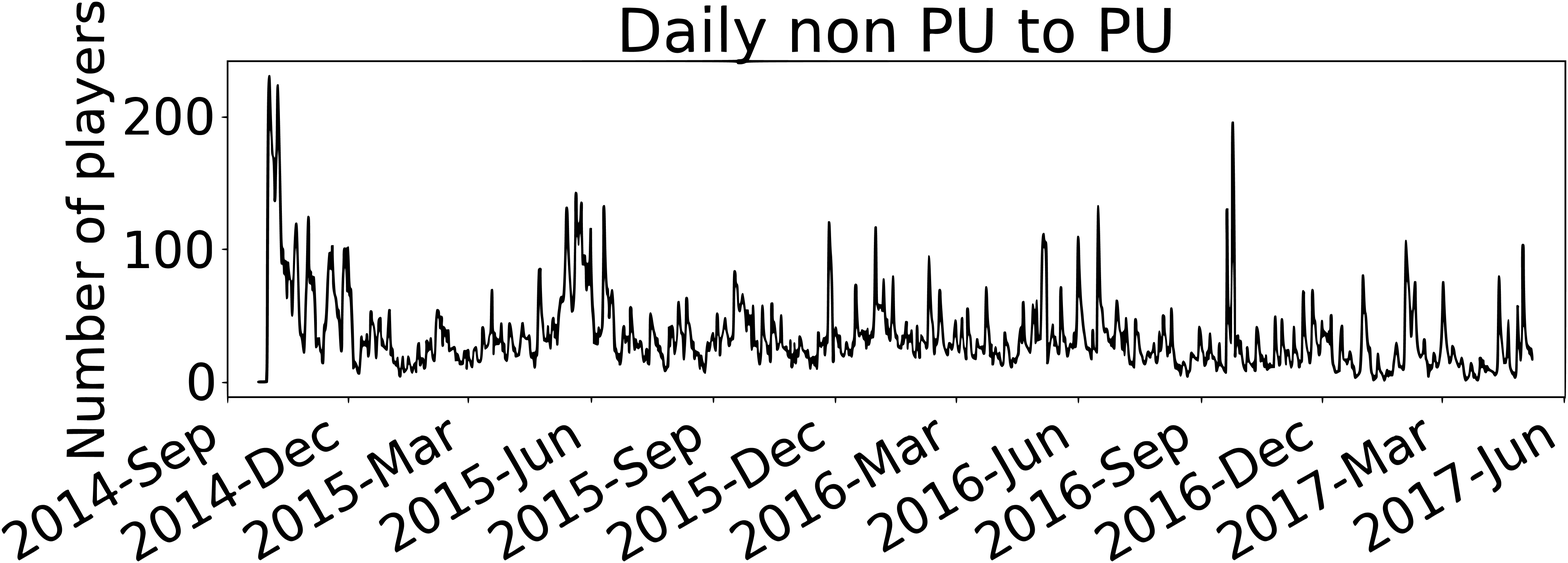}
\includegraphics[width=4cm]{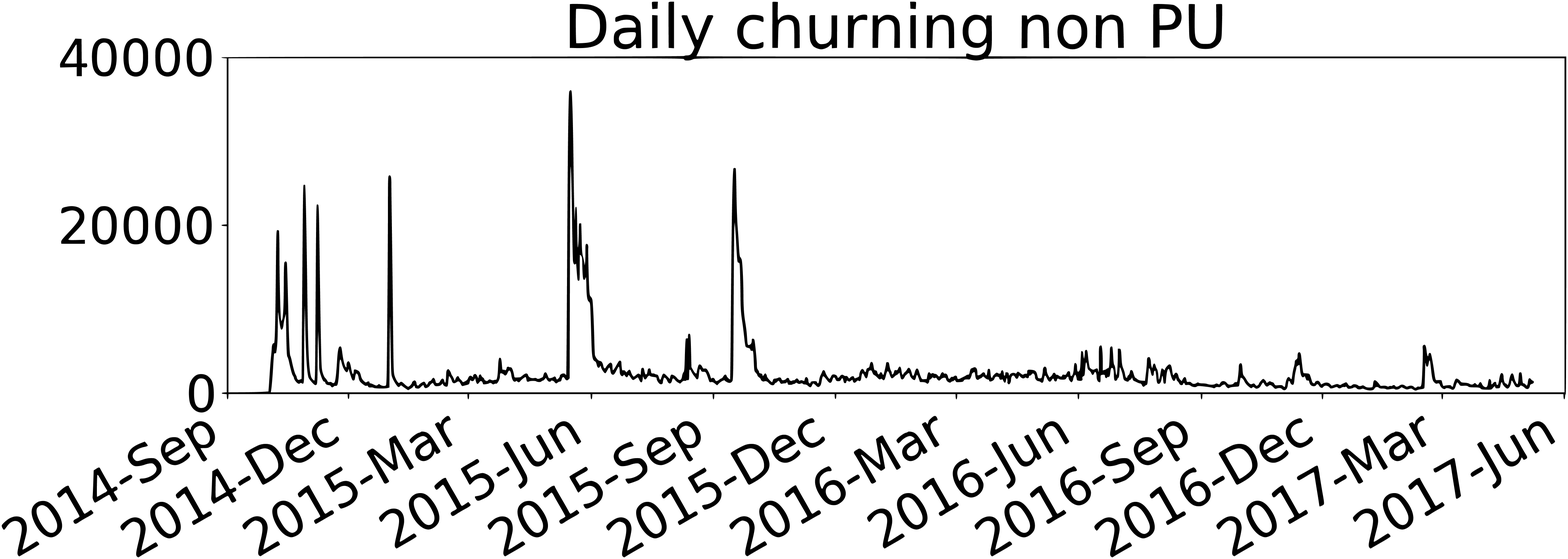}\\
\includegraphics[width=4cm]{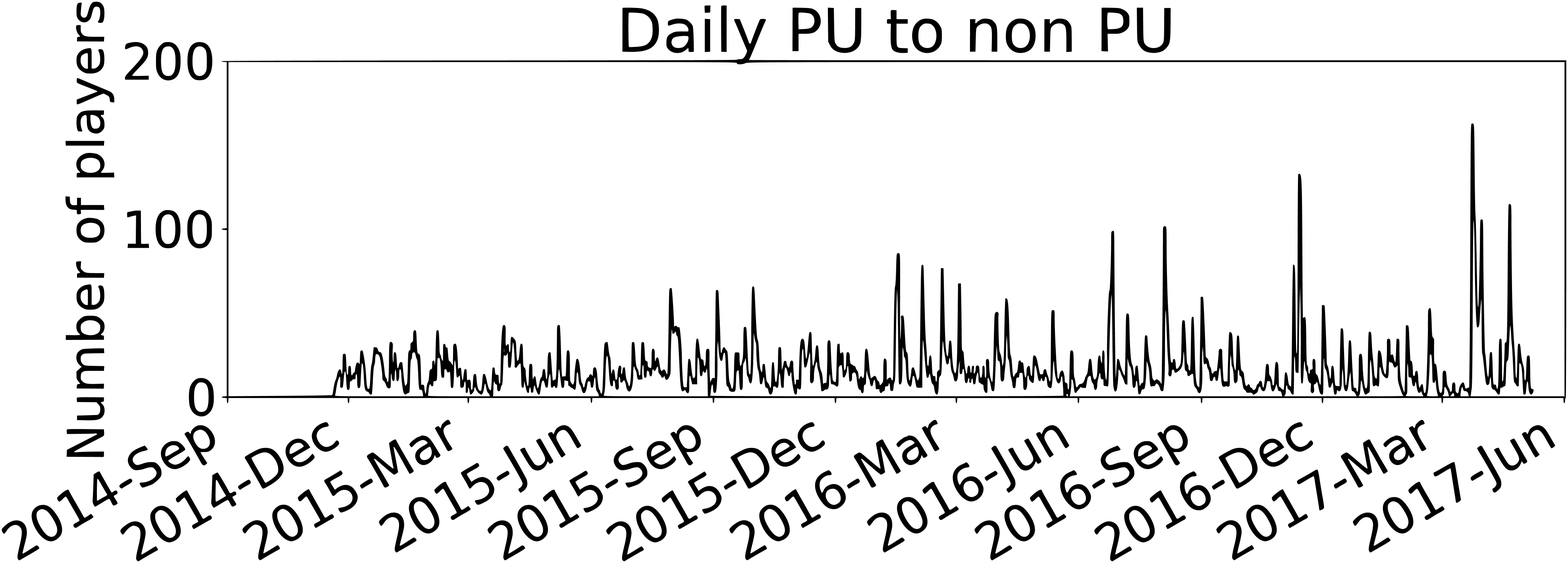}
\includegraphics[width=4cm]{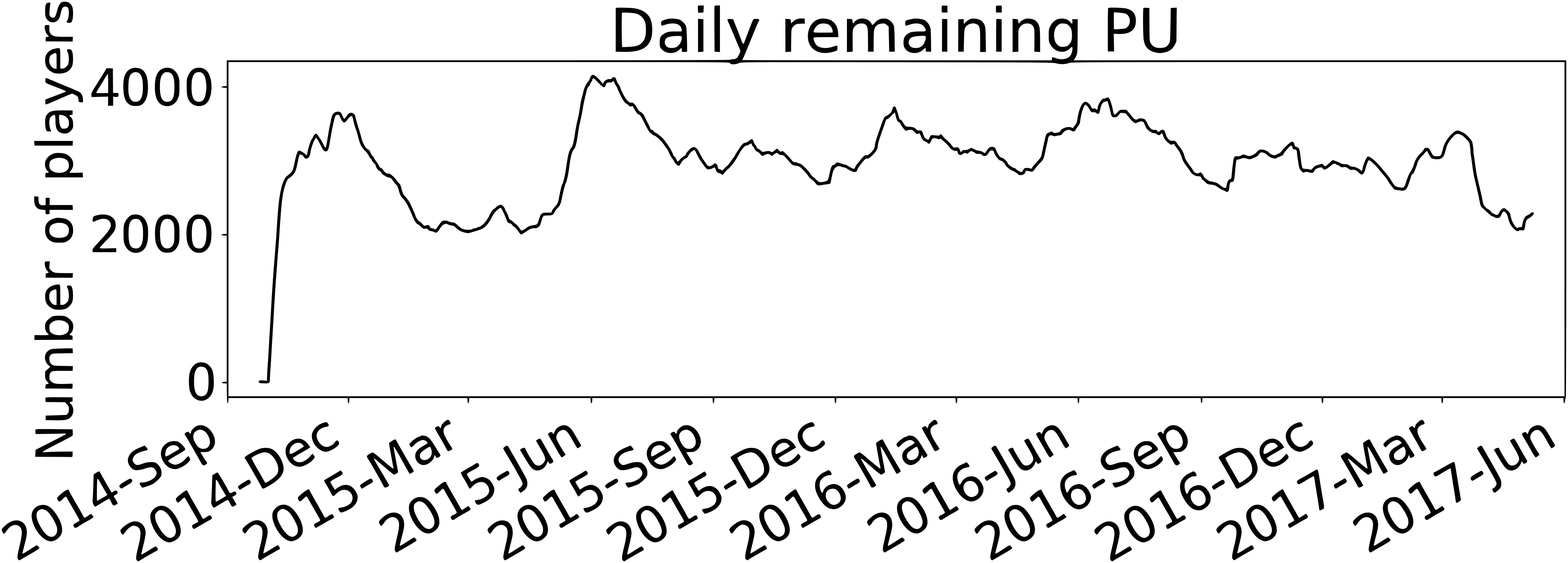}
\includegraphics[width=4cm]{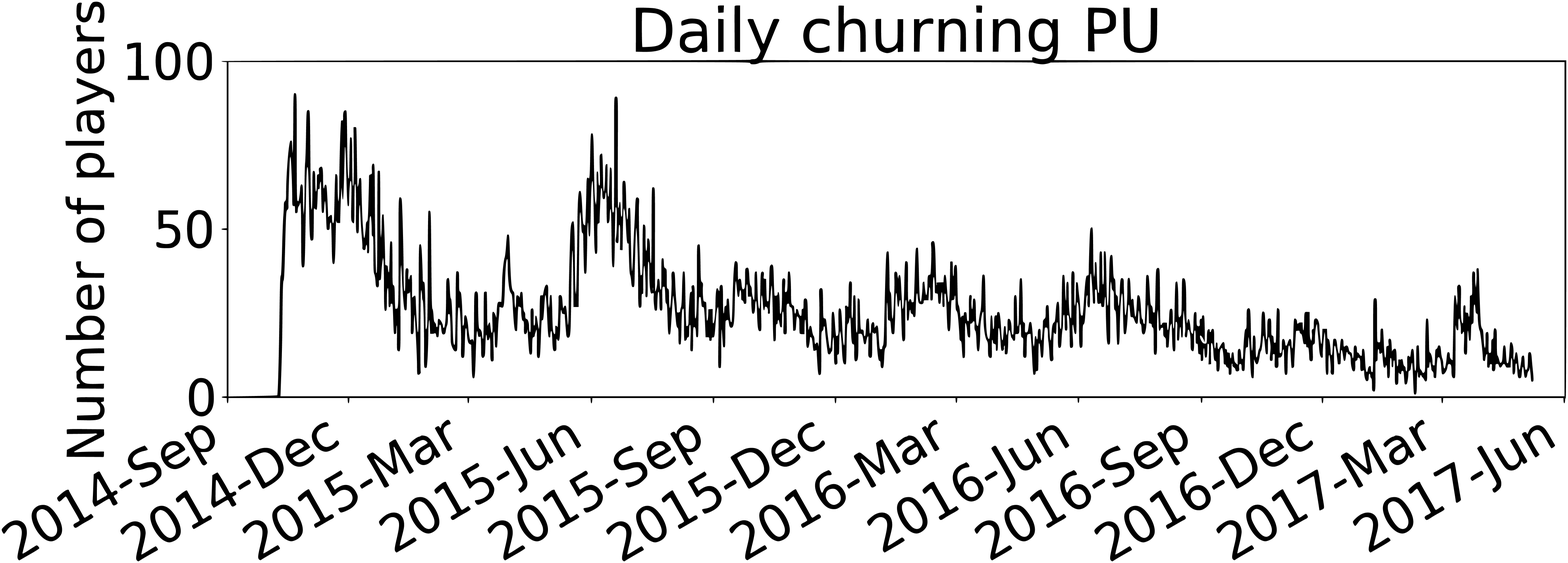}\\
\includegraphics[width=4cm]{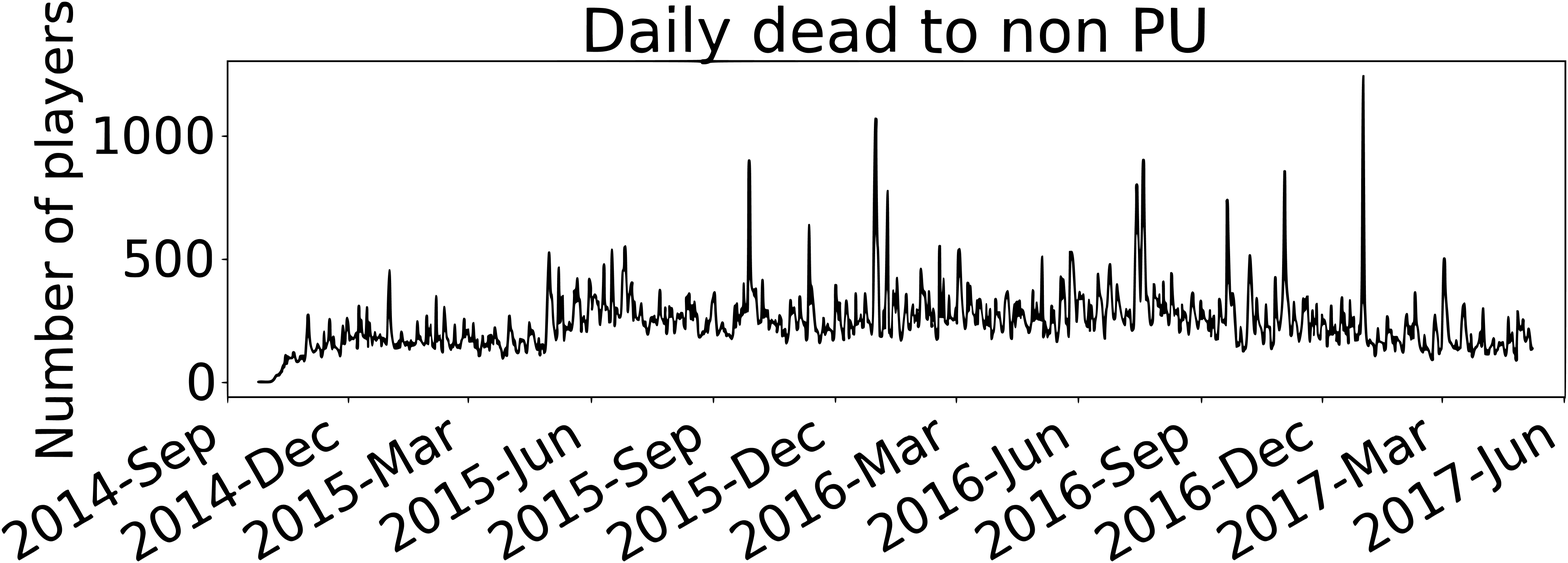}
\includegraphics[width=4cm]{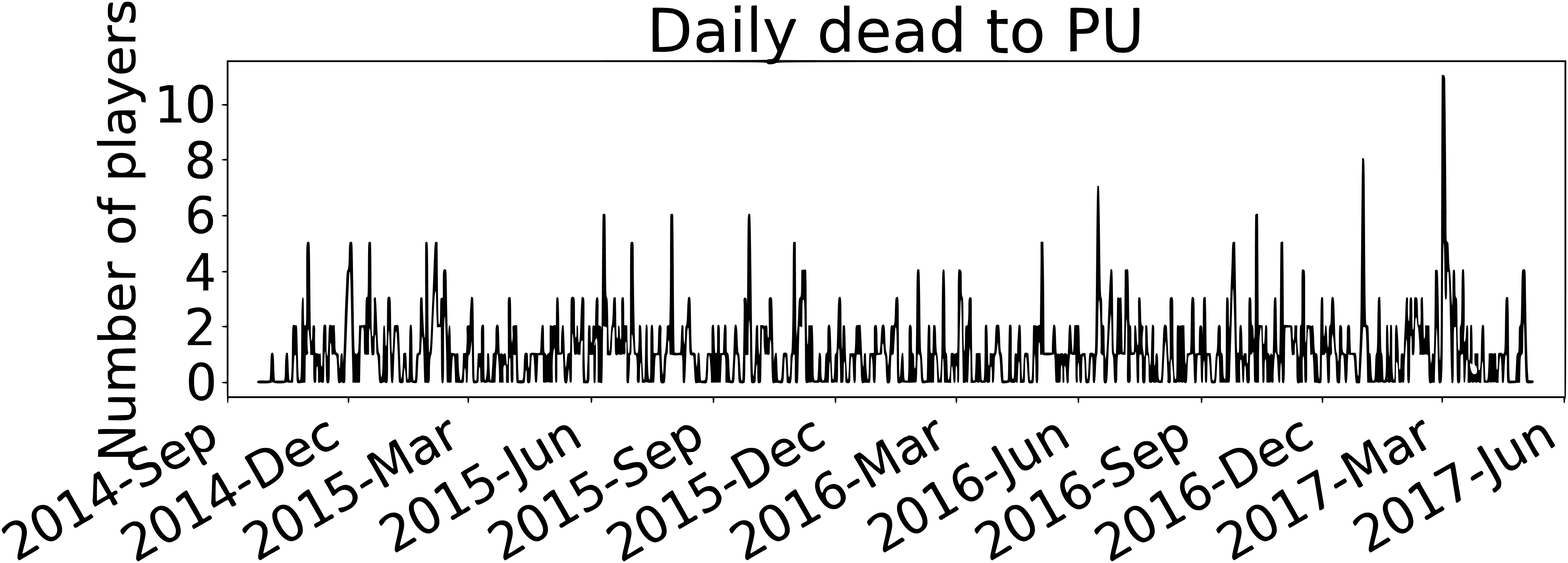}
\includegraphics[width=4cm]{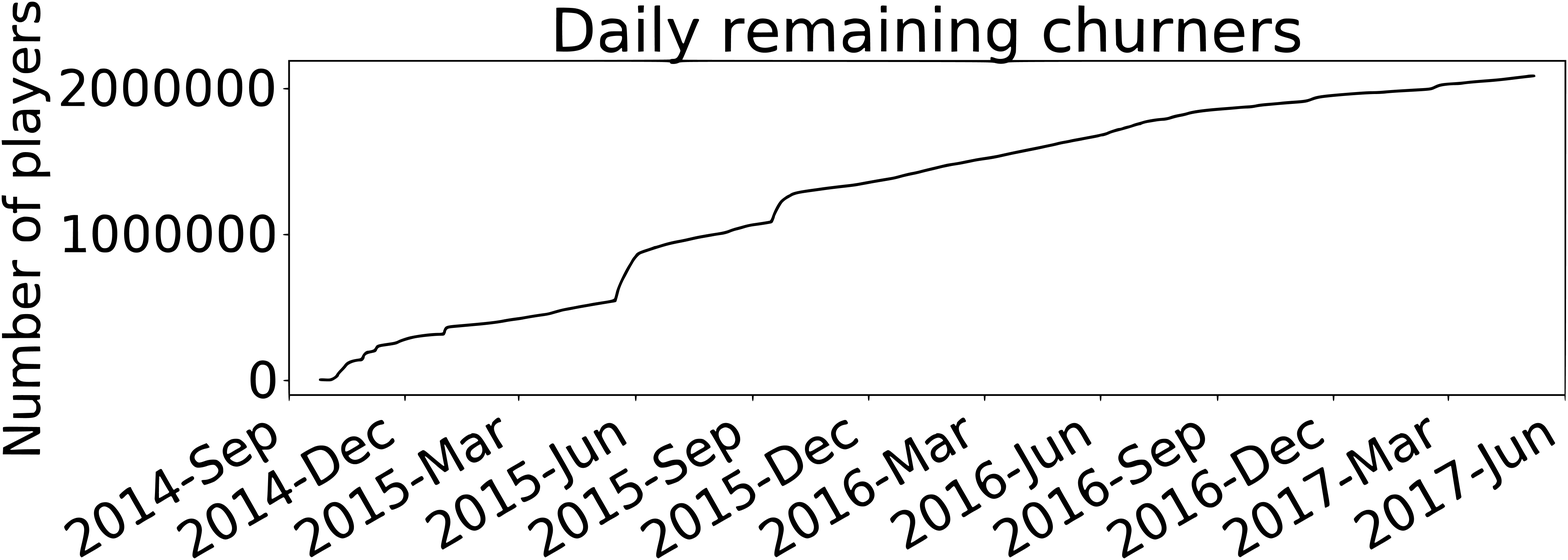}
\caption{Matrix of daily transitioning or remaining players between the three different segments considered. Top row concerns non-PUs, with plots for the number of them who are remaining non-PU (left), becoming PU (middle), or churning (right). Middle row refers to PUs who are: becoming non-PU (left), remaining PU (middle), or churning (right). Bottom row shows the number of churned players who are: becoming once again active non-PU (left), active PU (middle) or remaining inactive (right). }
\label{fig:trans}
\end{figure}

This information also allows for the construction of the (absolute) daily transition time series: \emph{new players} (users that log in for the first time that day), \emph{non-PU to PU} (players that purchase that day and had not made a purchase in the previous 50 days), PU to non-PU (players that purchased 51 days ago for the last time), \emph{non-PU to inactive} (players that have not purchased in the previous 50 days and logged into the game for the last time 10 days ago), \emph{PU to inactive} (players that have purchased in the previous 50 days and logged into the game for the last time 10 days ago), \emph{churned to PU} (players that log back into the game and make a purchase on that day after having been deemed churned) and \emph{churned to non PU} (inactive players that log back into the game and do not make a purchase on that day). Figure \ref{fig:trans} shows the matrix of transition and remaining series (in number of users).

With the population and transition series the conversion rates can be easily computed by dividing the daily transitions between the population of the group of origin on the previous day. These represent the daily probability of a user in a given group transitioning to a different one. In this work the transitions of churned players \emph{back to life} (players that become active again) are not considered, i.e, the
false churner probability will not be modeled. Both involve a small number of players and are of less interest than the other four conversion rates. The daily new user series is however also taken into account. This series is not only of utmost interest in itself, but is also crucial, as will be described, in the detection of the unknown marketing campaigns (as it is here where they will have the largest impact) and in discriminating these from promotion campaigns (that should have no measurable effect on it). 

%%%%%%%%%%%%%%%%%%%%%%%%%%%%%%%%%%%%%%%%%%%%%%%%%%%%%%%%%%%%%%%%%%%%%%%%%%%%%%%%%%%%%%%%%%
\section{Methodology}
\label{sec:meth}

Although the idea is to define a methodology that requires limited human intervention, the aim of this work is not to find a way of automatically producing forecasts. Human intervention is deemed necessary in acquiring qualitative knowledge of the systems and processes at play, which is one of the most important goals of this exercise. There must be however a fixed procedure guiding and limiting this intervention. This will allow for the use of this framework when defining more complicated segmentation landscapes. It also guarantees that different people will arrive to very similar or identical model definitions.

For each time series two different state space models -an ARIMA and a UC approach- both with covariates- are considered. The following five steps (described in some detail in the subsections below) describe the process followed for each time series to be modeled: (1) model selection; (2) selection of significant exogenous variables; (3) intervention definition; (4) model selection revisited; (5) forecasting and verification.

It is important to stress that the process described, though automatic to some extent, is still time consuming and relies heavily in human expertise. Though some steps could be taken to further simplify and automatize the process, this is inevitably going to be the case for the initial modeling phase and/or to use this approach to uncover missing information (marketing and promotion campaigns in this work). Once this phase is completed however, though some expertise would still be necessary periodically for model maintenance, the need for human intervention would be radically diminished. 

\subsection{Model selection}
\label{sec:model}

For the model definition and covariate selection and definition, all the historic data available is employed. The process begins with a general inspection of the series, its regular, weekly and monthly differences, and their correlograms. For the transition series, only additive models (non-transformed series) are analyzed. For the new users series, both the original and the log-transformed series are considered. The latter is selected as its variance is more stable. Regression parameters for explanatory variables in log-transformed series modeling (i.e. in multiplicative models) have a straight forward interpretation as elasticities, which is convenient to intuitively understand the parameters estimated. In the case of conversion rates (that already represent a fraction), parameters of additive models can also be understood in a straight forward manner as the increase in a day due to each unit of increase in the covariate series. This first inspection together with some basic stationarity tests decides in favour of a regular difference in the ARIMA case for all series. This leaves weekly structure to be accounted for, but taking a weekly difference yields a much higher anticorrelation and variance in the resulting series than a regular one in all cases.  

The starting training date (which will be different for the different time series) is decided upon. The behavior in the first days after launching the game is almost always very erratic and it is normally advisable to simply eliminate it. Further more, because of the nature of the launch, some series may start earlier than others (for example in this case purchases were not possible during the first days). In other cases the definition itself of the player segment (churn takes 9 days to be detected, purchase churn 50 days) accounts for these differences.

The two models -ARIMA and UC- to use as a base with which to explore the effects of other variables will be selected using \emph{brute exploration}. This means an estimation is run for the different possibilities of a selected subset of the model space (without linear regression to covariates) and the results compared in order to select the best option. The selection is done then through human intervention, but only considering the best 5 performing options according to the Akaike information criterion (AIC) \cite{akaike1974} after exploring a large amount of possibilities. Additionally, the Bayesian information criterion (BIC) \cite{schwarz1978}, the Hannan-Quinn information criterion (HQIC) \cite{hqic}, residual\footnote{\emph{Residuals} refer to the unexplained part of the series after modeling. They should correspond to the random noise term described in section \ref{sec:ssm}.} variance, independence and normality, and parameter significance are taken into account to select one option, always favouring less parameters for similar performance. Ljung-Box \cite{ljung-box} and Jarque-Bera \cite{jarque-bera} tests are used to assess independence and normality of the residuals respectively. Parameter significance is evaluated using Z-scores \cite{kreyszig}, with parameters with associated p-values under 0.1 considered significant. If less human expertise is available and/or the time for the model definition phase wants to be reduced, the best AIC performing option could be selected.

In the ARIMA case estimations are run for all possible combinations of weekly and regular ARMA polynomials of order up to 5 (in both AR and MA). This is a very extensive exploration designed to minimize the expertise and time devoted to the preliminary phase of time series inspection, and to make the process as automatic as possible. Given that the different combinations can be run in parallel and the ARIMA estimation is not computationally expensive, this is in general a reasonable approach. The use of higher than order two ARMA polynomials is however rarely justified, so the parameter space to be explored could be bounded to lower orders. Carefully analyzing the differently differenced series and their correlograms would also allow for a selection of only a few different models to try, and this would also be a valid option.

Interestingly, although there is a very clear weekly structure in at least the PU churn and purchase churn series (as shown by the significant correlations in the correlograms for lags 7 and some of its multipliers), this analysis favours in all cases models with no seasonal ARIMA. Weekly effects will be therefore accounted for using day of the week exogenous variables as described in section \ref{sec:exog}.

In the UC case, the use (or not) of a cyclic term and the use (or not) of weekly and monthly seasonality is explored. Monthly seasonality (as the preliminary analysis suggested) is rejected in all cases. In regards to weekly seasonality, its use is deemed favourable in all cases (and it will be used instead of the day of the week covariates employed for the ARIMA models). For the level-trend, different options are also explored: no trend, fixed intercept (deterministic constant), local level (random walk), fixed slope (deterministic trend), local level with deterministic trend (random walk with drift), local linear trend and smooth trend (integrated random walk). In all cases, the local level type of trend is the best option. As it would be expected, other options with more degrees of freedom have an additional reduction of the residual variance, but have notably worse information scores. Here again a more careful initial exploratory analysis could limit the number of models to be tried, but this hardly seems justified as UC models are even less computationally expensive than ARIMA ones, and the parameter space explored is in any case smaller.

\subsection{Exogenous variable selection}
\label{sec:exog}

With the information available, the following explanatory time series to be used as covariates are built:

\begin{itemize}
    \item \emph{Day of week}: Effects for each day of the week (one variable per day of the week which is 1 that day and 0 elsewhere).
    \item \emph{Calendar effects}: First of month, last of month, first of year and last of year effects (estimated separately).
    \item \emph{Holidays}: National holidays and school holidays effects are considered separately (with effect estimated jointly for all days in each of these two groups).
    \item \emph{In-game events}: All events of the same type and with the same event scale are considered jointly. Out of each type of event and event scale two inputs are built: event on/off (covariate is 1 when there is an event of that type and 0 elsewhere) and event start (covariate is 1 when an event of that type is beginning on that day and 0 elsewhere).
    \item \emph{Number of in-game events}: Besides inputs for each event type and event scale combination, two additional inputs with values number of events going on that day and number of events starting on that day are also considered.
\end{itemize}

Additionally, interventions will be defined for each of the time series as described in section \ref{sec:int}. These interventions will be tried as exogeneous variables not only for the series for which they were detected but also for the rest. This means that the methodology described will be repeated twice for all of the series to ensure that any and all interventions are tested for all series.

To decide which of the variables to use with each time series, the available covariates are added progressively in groups, each time discarding those with parameters that are not estimated to be significant. Significance is evaluated using Z-scores \cite{kreyszig}, and parameters with associated p-value larger than 0.1 are rejected. The grouping and order in which explanatory variables are tried is that of the enumeration above in the \emph{first round}. When a group is made up of more than ten inputs (i.e. for in-game event covariates), these are tried in groups of ten. The handling of interventions is very similar and is described in more detail in section \ref{sec:int} (as well as the order used in trying all covariates in the \emph{second round}). After covariates from all groups including interventions have been selected in this way, all variables that have been left out are then tried again one by one to make sure that they are still not significant with the final configuration.

Naturally, for churn probability series exogenous variables are always introduced with a delay corresponding to (regular or purchase) churn definition (10 and 51 days respectively for this game).

\subsection{Interventions}
\label{sec:int}

Interventions are exogenous variables defined \emph{adhoc} after analyzing the residuals of a previous covariate configuration. They should capture the effect of the most important marketing campaigns aimed at new user acquisition, as well as promotion campaigns aimed at conversion to PU or enhanced spending of PUs, of which, as it has been already noted, no information is available (except for the fact that they did exist and had significant impact). The detection and classification of these campaigns is one of the main goals of this work. 

Marketing and promotion campaigns (or other effects of unknown origin with significant impact in the transition rates) are expected to leave very large residuals in the absence of these interventions. The procedure to construct them will be to start with the day with largest deviation in the residuals. Human inspection is needed to decide the exact shape of the intervention. If, for example, a very large positive residual is followed by a large negative one several days later for the ARIMA model (which always uses a regular difference), a campaign will be assumed to have run starting on the day with large positive residual and ending the day before the negative one. The model is then reestimated with the intervention designed to capture the effect seen in the original series and the residuals. If the paramater is significant, the variance is reduced, and if Jarque-Bera normality test of the residuals yields a better score, the intervention is kept, and a new intervention is included for the next largest deviation in the new residuals. This process is repeated until adding new interventions makes the residual less and not more normal.

As described in the previous section \ref{sec:exog}, interventions \emph{discovered} for any of the series will also be tried on the other ones. Depending on the type of effect and on which series is estimated as significant, interventions are classified as:

\begin{itemize}
    \item \emph{Marketing interventions}: These are outliers which look like they could be a result of marketing (out of game) or new user acquisition campaigns. They should have a strong positive impact in the new user series. Typically, they will also have a positive effect on churn transitions (after 9 days), unless they have been particularly good at getting to the right target (i.e., people that have actually kept playing after the first day they tried the game).
    This effect is expected to be larger for non-PU churners (as people who try the game and rapidly move to make their first purchase are more likely to be really interested in the game and thus continue playing). They could also possibly have a (limited) positive effect in the conversion to PU series, as it could also encourage spending in people that are already playing but are exposed to the campaign. In this case, they could also have some impact in purchase churn probability 50 days afterwards.
    \item \emph{Promotion interventions}: These should reflect promotions offered to players, and are therefore mainly characterized by having a strong positive impact on the probability of conversion to PU. They will also typically have a measurable effect on the purchase churn probability 50 days later, and the difference of impact in both series for different campaigns will help detect which have been more useful in generating long term conversion. They could have some minimal impact on churn probabilities if they have been particularly bad (if players are spammed with notifications for example). Never should there be any mensurable effect in new users, as these are promotions that are only available for already existing players.
    \item \emph{Unknown interventions}: Outliers in a different direction from what would be expected due to marketing or promotion campaigns. These could be related to other relevant missing information such as server problems, buggy releases, changes in the game dynamics or content, etc. 
\end{itemize}

Interventions are tried with all series in the same way as the rest of covariates and considering the three groups listed above (and grouped in tens when needed). The series were modeled in the following order: (1) new users (in an attempt to discover as many new user acquisition or marketing campaigns as possible); (2) conversion to PU (in search for missing important promotion campaigns); (3) non-PU churn (for further marketing campaign detection, as promotions should have none to very little impact), (4) PU churn and (5) purchase churn (where further promotions can be detected).

After this first round all additional covariates in form of interventions are assumed to have been detected, and the exogenous variable selection process is then repeated starting from scratch for all series. In this \emph{second round}, taking into account the nature of each of the series modeled and of the impact the different interventions and in-game events are expected to have, the order in which the different groups are tried varies slightly, with marketing interventions being tried before in-game events and the rest of interventions for new users and before the rest of interventions for both churn series; and promotion interventions being tried before the rest of interventions for conversion to PU and purchase churn. Unknown interventions are tried last in all cases.

This process yielded a plausible campaign scenario, as will be described in section \ref{sec:results}, when working with ARIMA models. The process was more cumbersome and less effective when dealing with UC models, as they have several components and noise terms that can better capture sudden rises and drops in the series without the need of interventions. It was finally decided to carry out the intervention definition process with the ARIMA models only and then use these for both ARIMA and UC on what has been described as \emph{second round}.

This is the most time consuming part of the process, and the one where expert human intervention is more critical. This is however unavoidable if there is missing information (marketing and promotion campaigns in our case) that should be unveiled in the process. If finding out the more plausible particular scenario is not a priority, a fully automatic simplified approach could be followed. Namely, the largest outliers in the residuals could be corrected using a single additional variable for that day. Analogously to the process described above, these would be then accepted or rejected depending on whether its associated parameter is estimated as significant or not, and whether it improves or not the normality tests. This process would be repeated automatically until the introduced variable is not significant or the normality test is degraded. This would correct outliers, yielding a more consistent model and preventing it from learning from atypical values. The anomalous realizations of the series will remain however unaccounted for. In addition, a threshold could be introduced below which the normality test would be considered valid and the iterative method interrupted, as specially without human control, it is important to avoid overfitting the model. Even in cases where there is not known missing relevant information, it is convenient to follow such a process in order to ensure normality of the residuals, as the models would not be formally valid otherwise, and failure to correct unusual behaviour would result in underfitting.

\subsection{Model selection revisited}

After having a final set of exogenous explanatory variables with which to proceed, the model space is again revisited and the best AIC scoring options analyzed in some detail again. Although in some cases there were slight changes from the originally selected model definition described in \ref{sec:model}, the main findings described there hold. Namely, in all cases ARIMA models with a regular difference and without weekly polynomial and UC models with local level outperform the others.

\subsection{Forecasting and verification}
\label{sec:fc}

Finally, after selecting the best model definition with the available data, daily forecasts are run for all of 2016 and what is available of 2017 for verification. Replicating a possible production setup, new daily forecasts are run for each month using data until the last day of the previous month to train the model. However, compared to a real production setup, the current model and exogenous variable selection made use of more data (all historic data available). Nevertheless, it still only uses for each training that will produce the forecasts the data that would have been available at the time. Besides, as the interventions used are always local, information on future interventions (planned marketing and promotion campaigns, or other unexpected events such as buggy updates or server failures) will not be available for the models, accounting for large forecast errors due to missing relevant information. Monthly Mean Absolute Error (MAE) and Root Mean Square Error (RMSE) are then examined and compared for the different models and series to assess forecast accuracy. Note the aim of computing these validation metrics is that of comparing the performance of both models, and of each model for different months, rather than assessing the overall goodnes of any of the two models, as there is no appropriate baseline to which to compare them.

%%%%%%%%%%%%%%%%%%%%%%%%%%%%%%%%%%%%%%%%%%%%%%%%%%%%%%%%%%%%%%%%%%%%%%%%%%%%%%%%%%%%%%%%%%
\section{Results}
\label{sec:results}

\begin{figure}
\centering
\includegraphics[width=11cm]{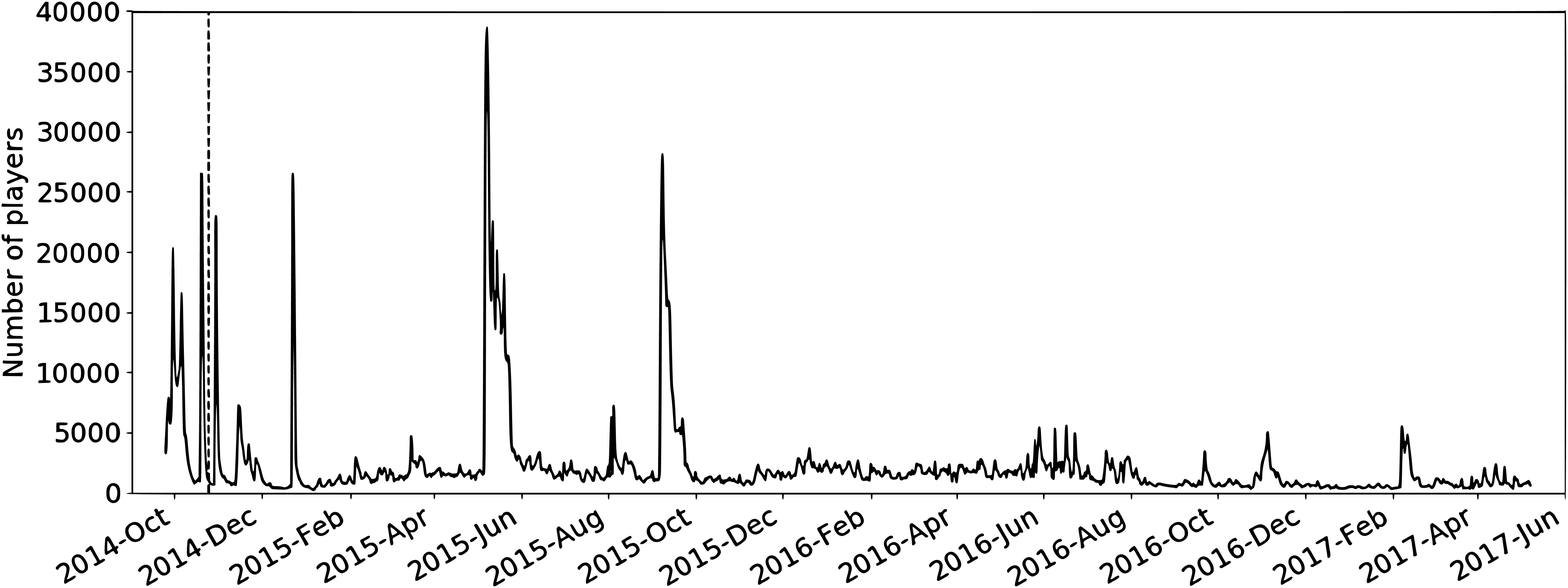}\\
\includegraphics[width=11cm]{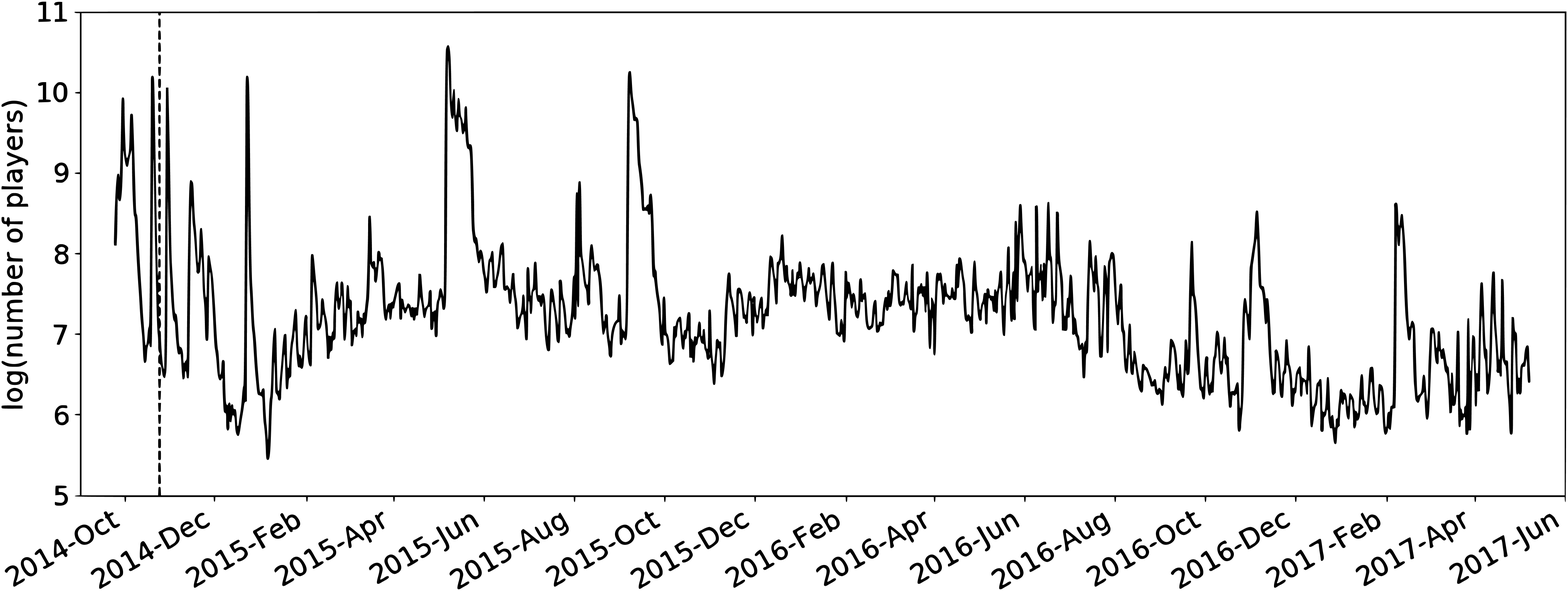}\\
\includegraphics[width=11cm]{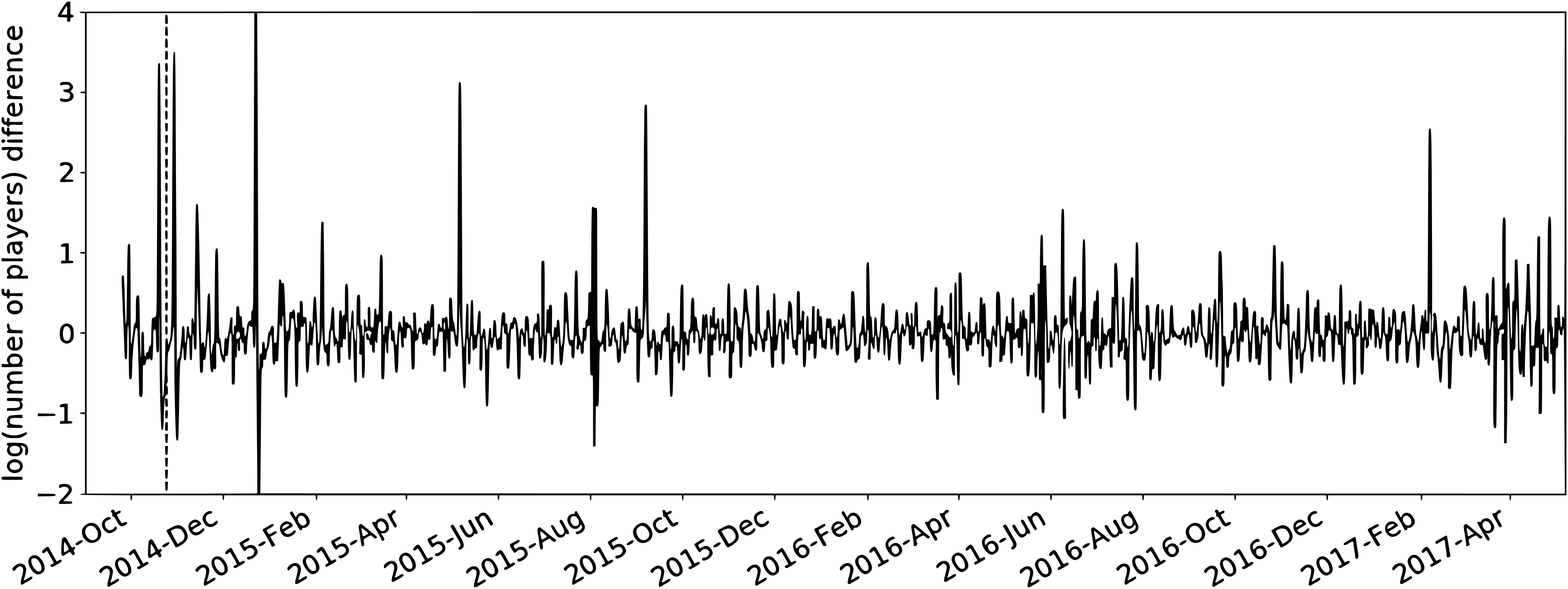}\\
\includegraphics[width=5.5cm]{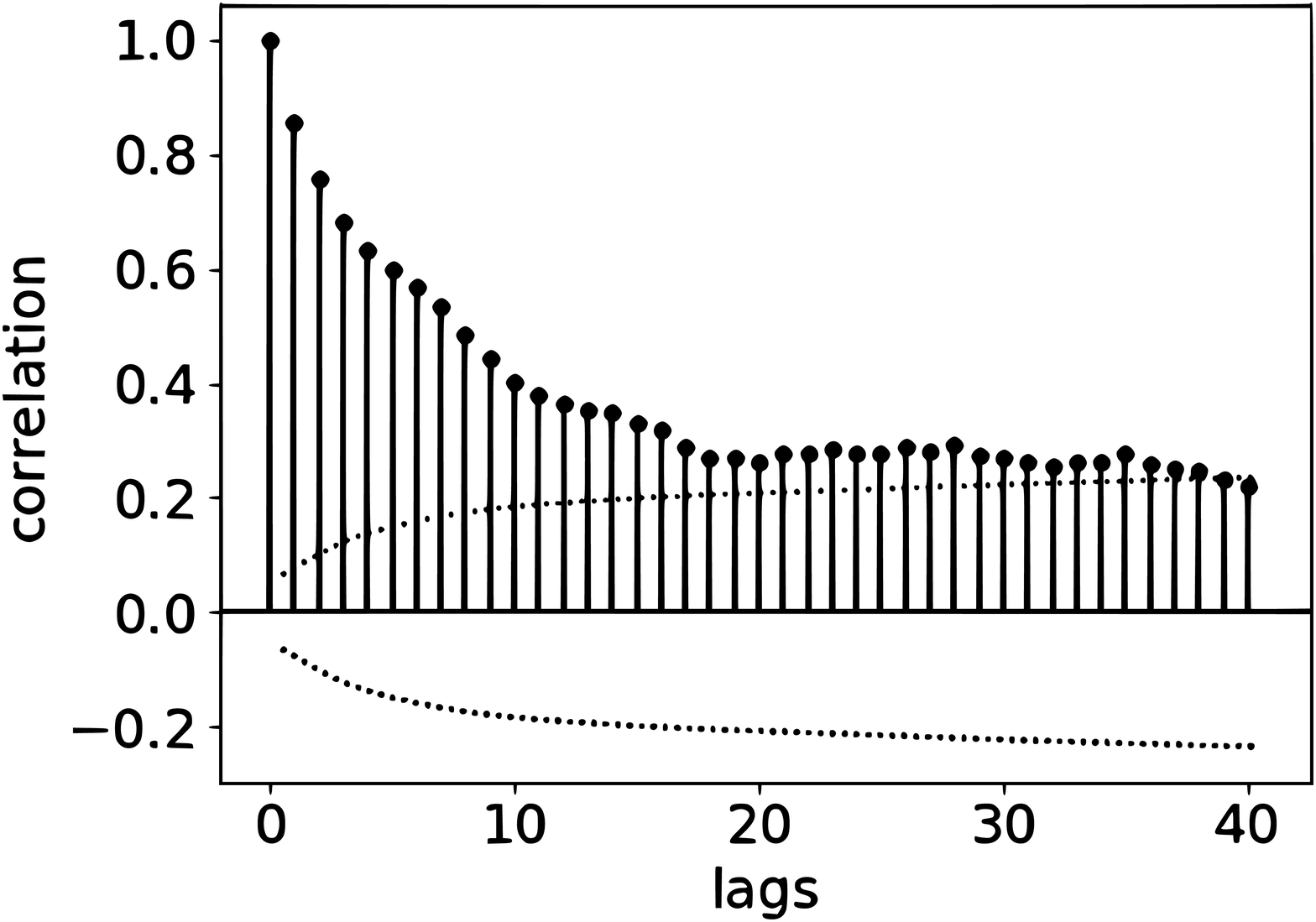}
\includegraphics[width=5.5cm]{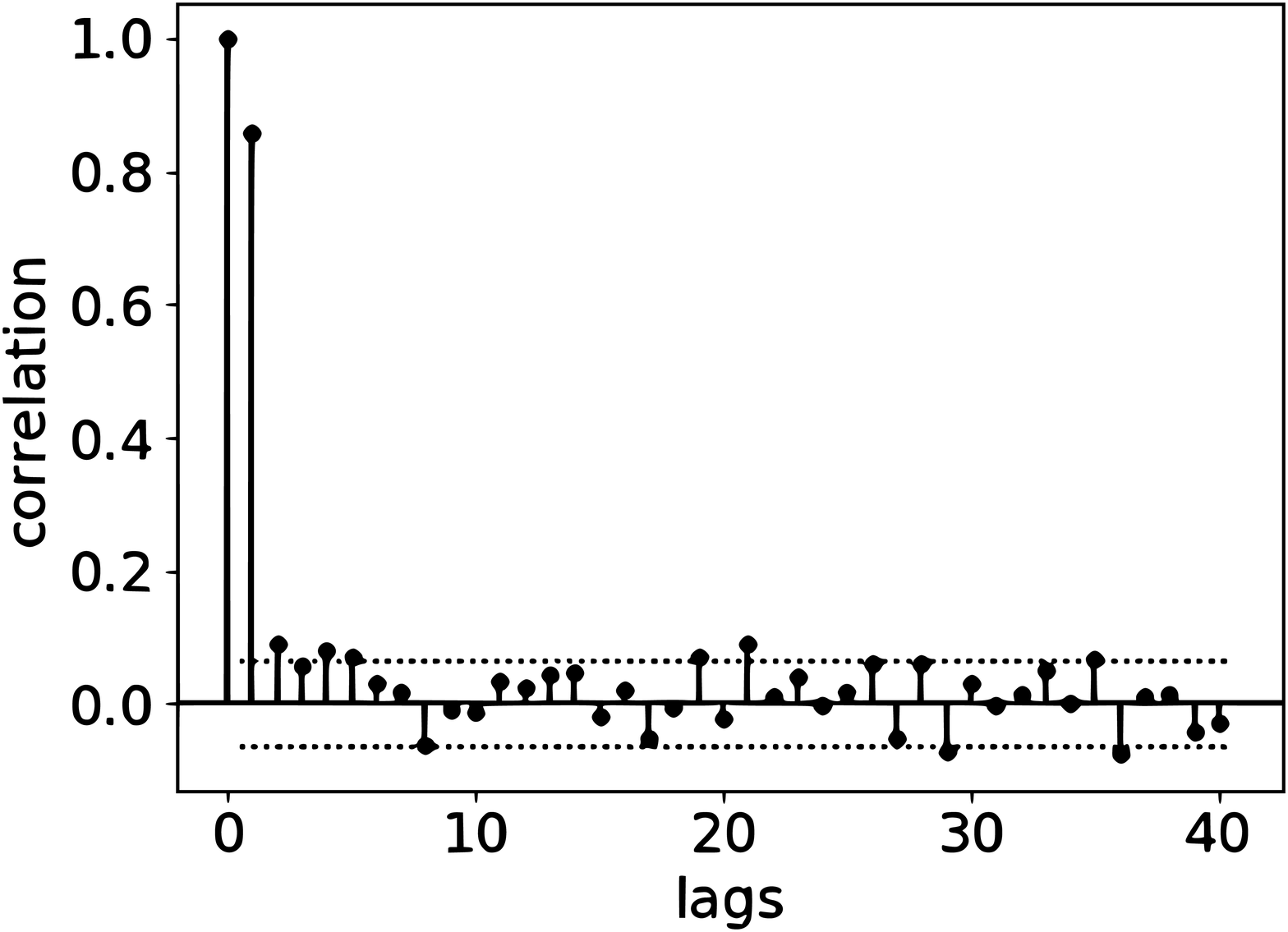}
\caption{Daily new users original series (top), log-transformed (second row), difference of the log-transformed (third row), ACF (bottom left) and PACF (bottom right). Start of the training period is marked with a dashed line and corresponds to October 10, 2014. The dotted lines in the ACF and PACF define the significance region (for values larger than the positive line or smaller that the negative one) with 95\% confidence.}
\label{fig:new}
\end{figure}

The daily number of new users (players who log into the game for the first time) is shown in Figure \ref{fig:new}'s top plot, with a dashed line marking what has been taken as starting day for the training, October 10, 2014 for this series. The ARIMA used was (2, 1, 1) and the local level model with weekly seasonality used a longer cycle periodicity too. In both cases the series was log-transformed (all prediction error measures given refer to the untransformed forecast though). The log-transformed series together with its regular difference are Figure \ref{fig:new}'s second and third plots from the top respectively (again with a dashed line marking the starting date for the training). The two bottom figures correspond to the Autocorrelation Function (ACF, left) and Partial Autocorrelation Function (PACF, right) of the log-transformed. For this and all other series, the ACF and PACF were computed leaving out the period before the training starting date. In the correlograms, regions outside the area delimited by the dotted lines correspond to significant correlation values with 95\% confidence.

\begin{figure}
\centering
\includegraphics[width=11cm]{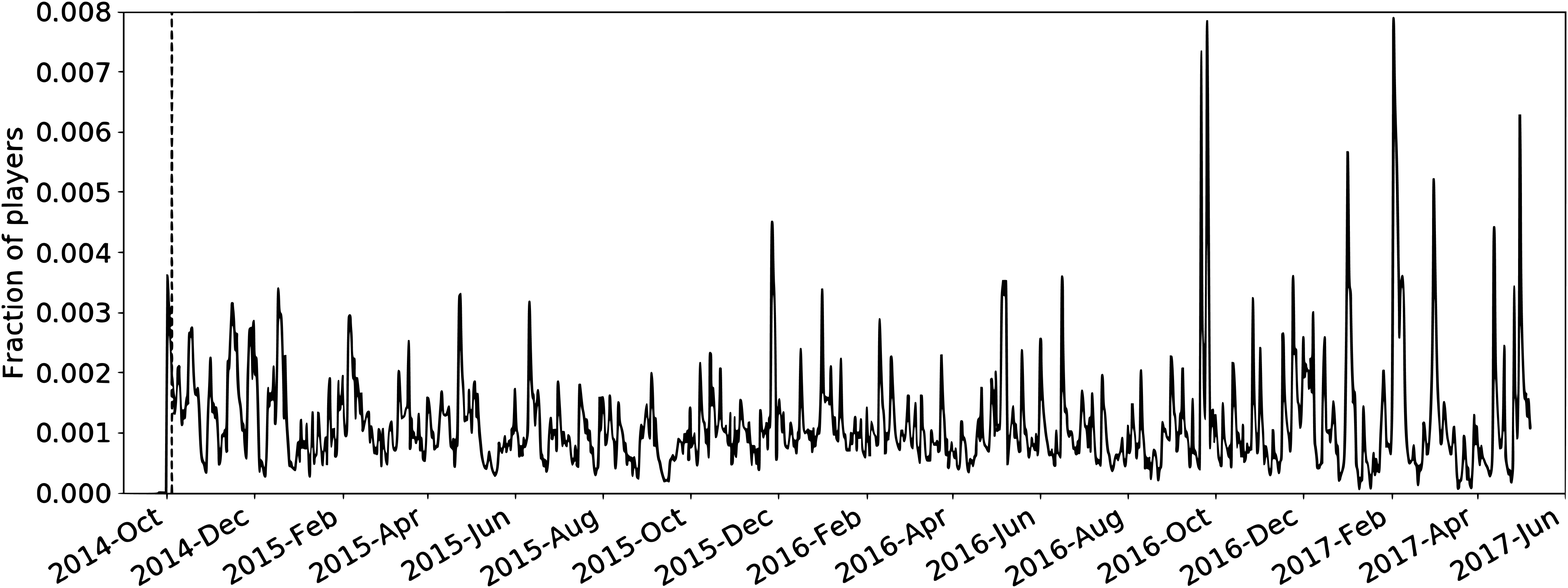}\\
\includegraphics[width=11cm]{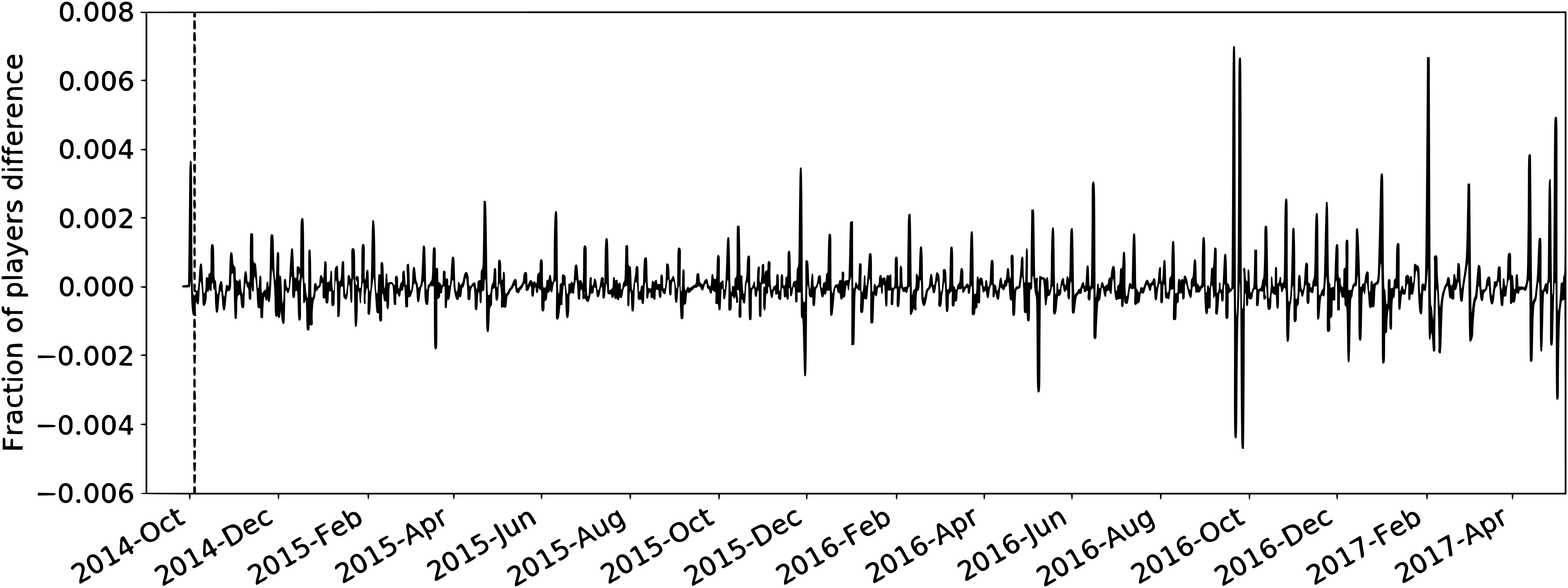}\\
\includegraphics[width=5.5cm]{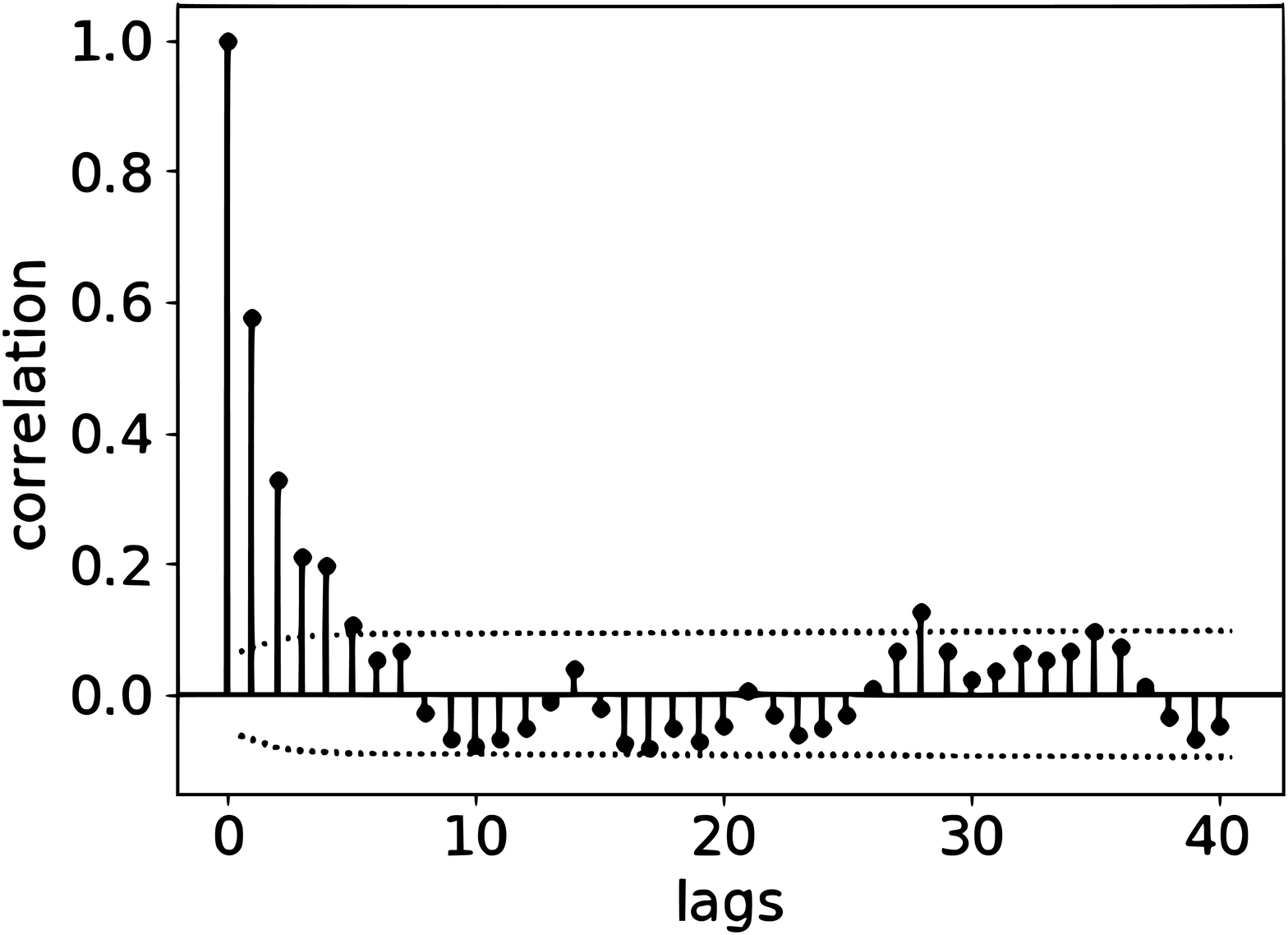}
\includegraphics[width=5.5cm]{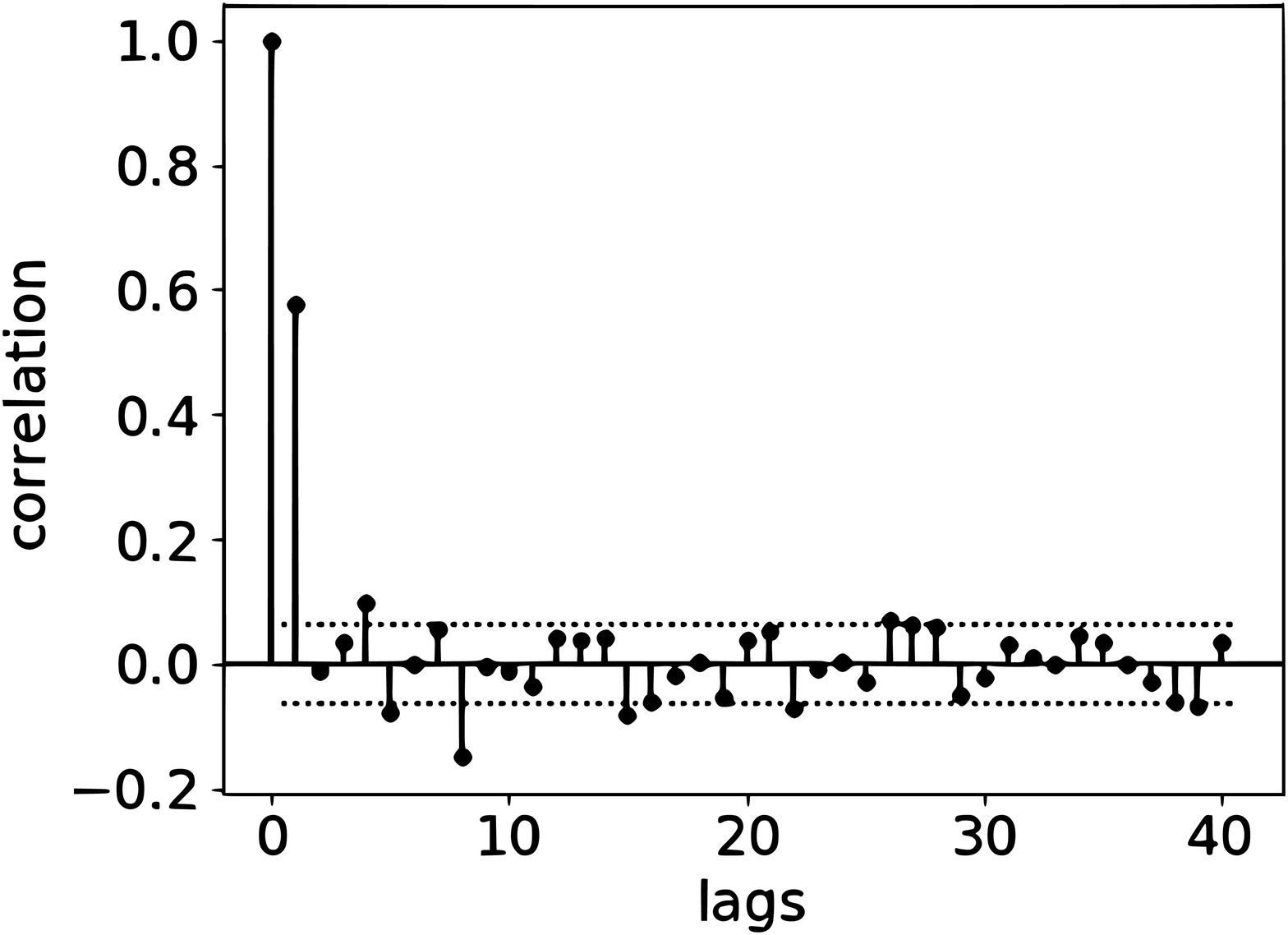}
\caption{Daily non-PU to PU conversion original series (top), its regular difference (middle), ACF (bottom left) and PACF (bottom right). Start of the training period is marked with a dashed line and corresponds to October 5, 2014. The dotted lines in the ACF and PACF define the significance region (for values larger than the positive line or smaller that the negative one) with 95\% confidence.}
\label{fig:pnp}
\end{figure}

\begin{figure}
\centering
\includegraphics[width=11cm]{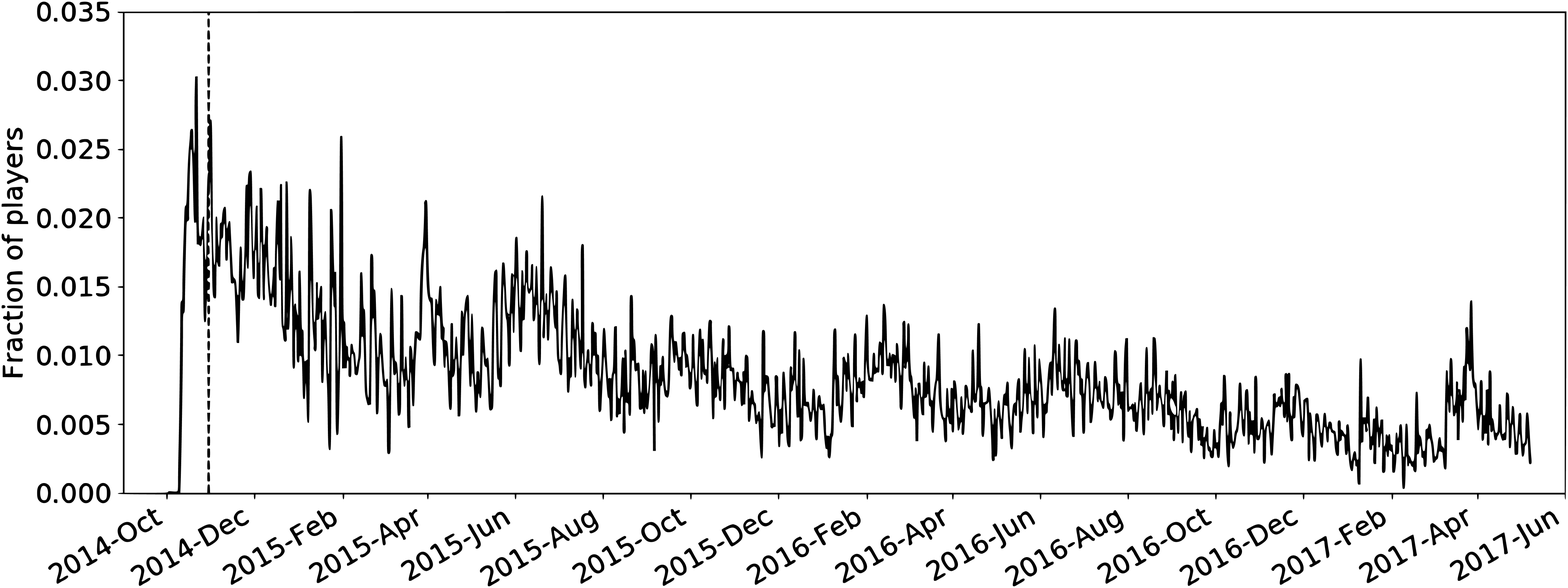}\\
\includegraphics[width=11cm]{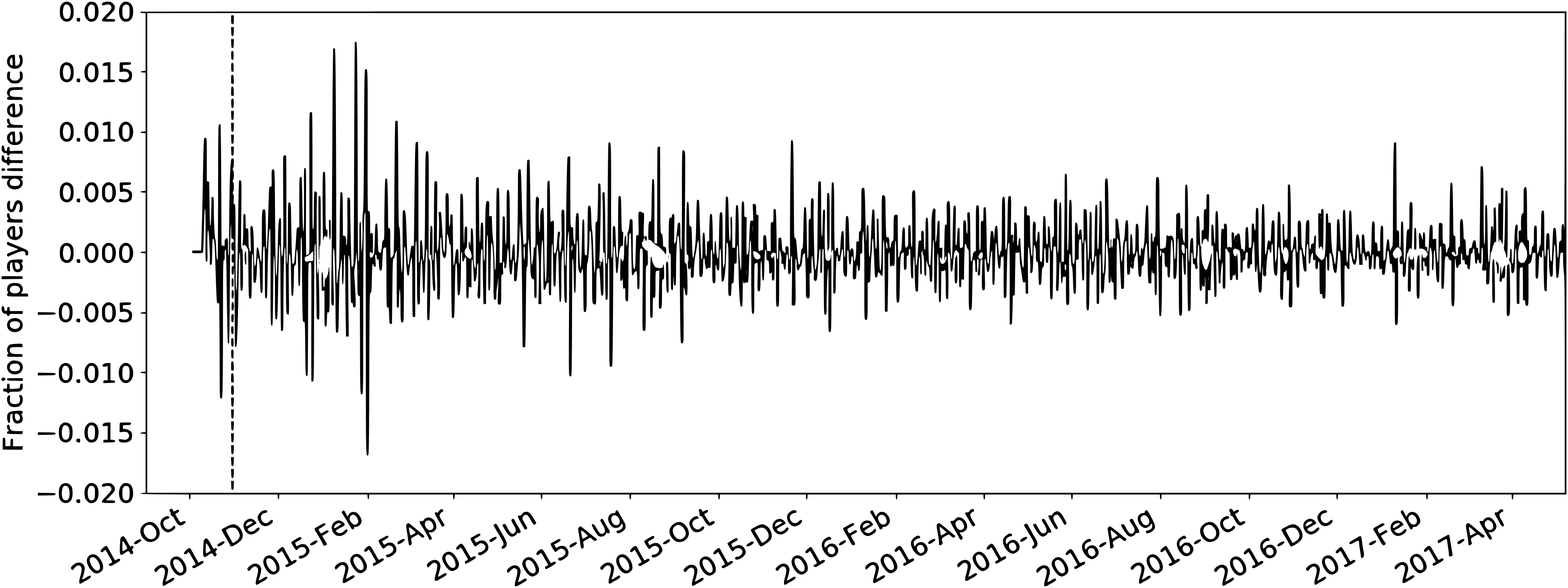}\\
\includegraphics[width=5.5cm]{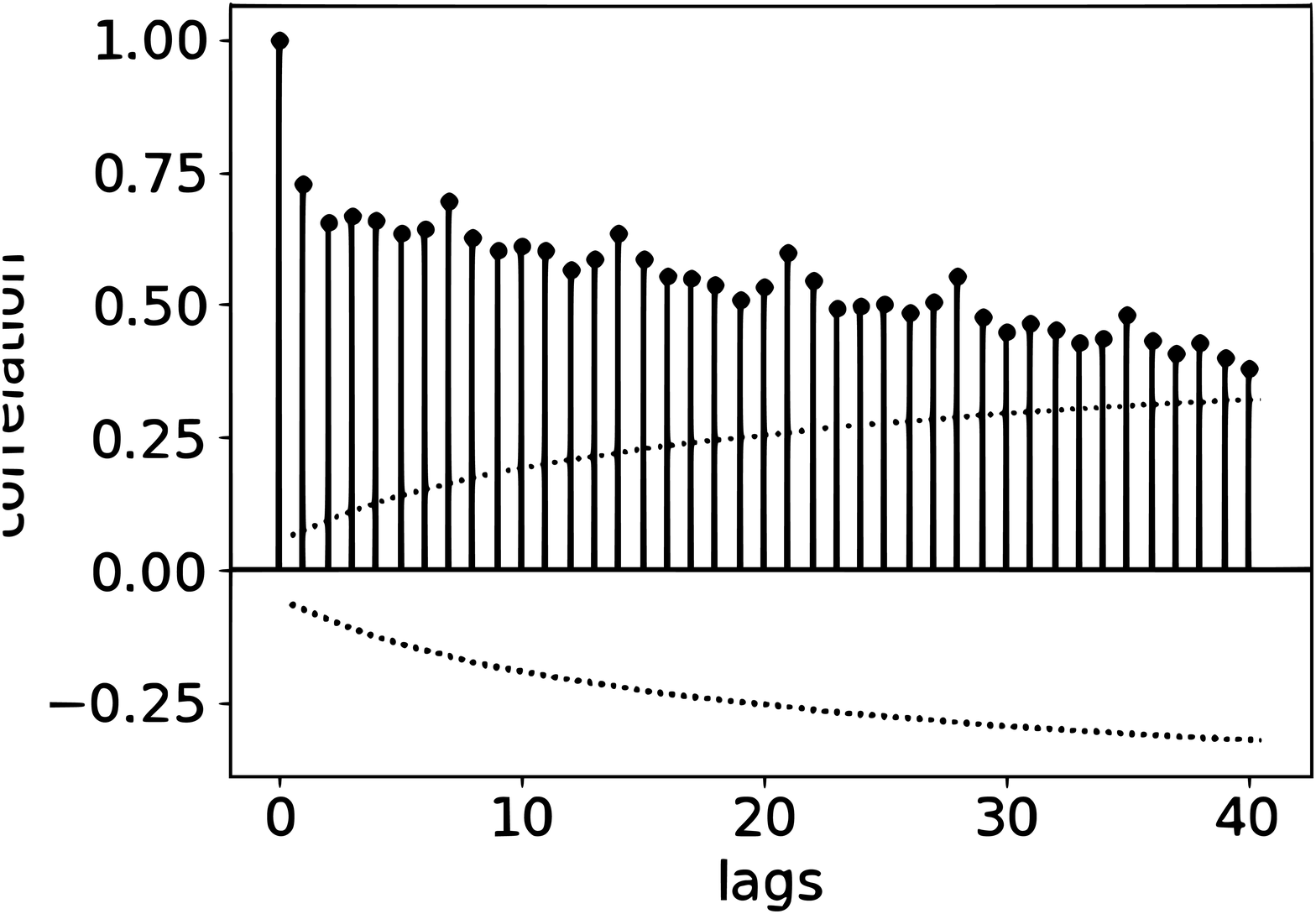}
\includegraphics[width=5.5cm]{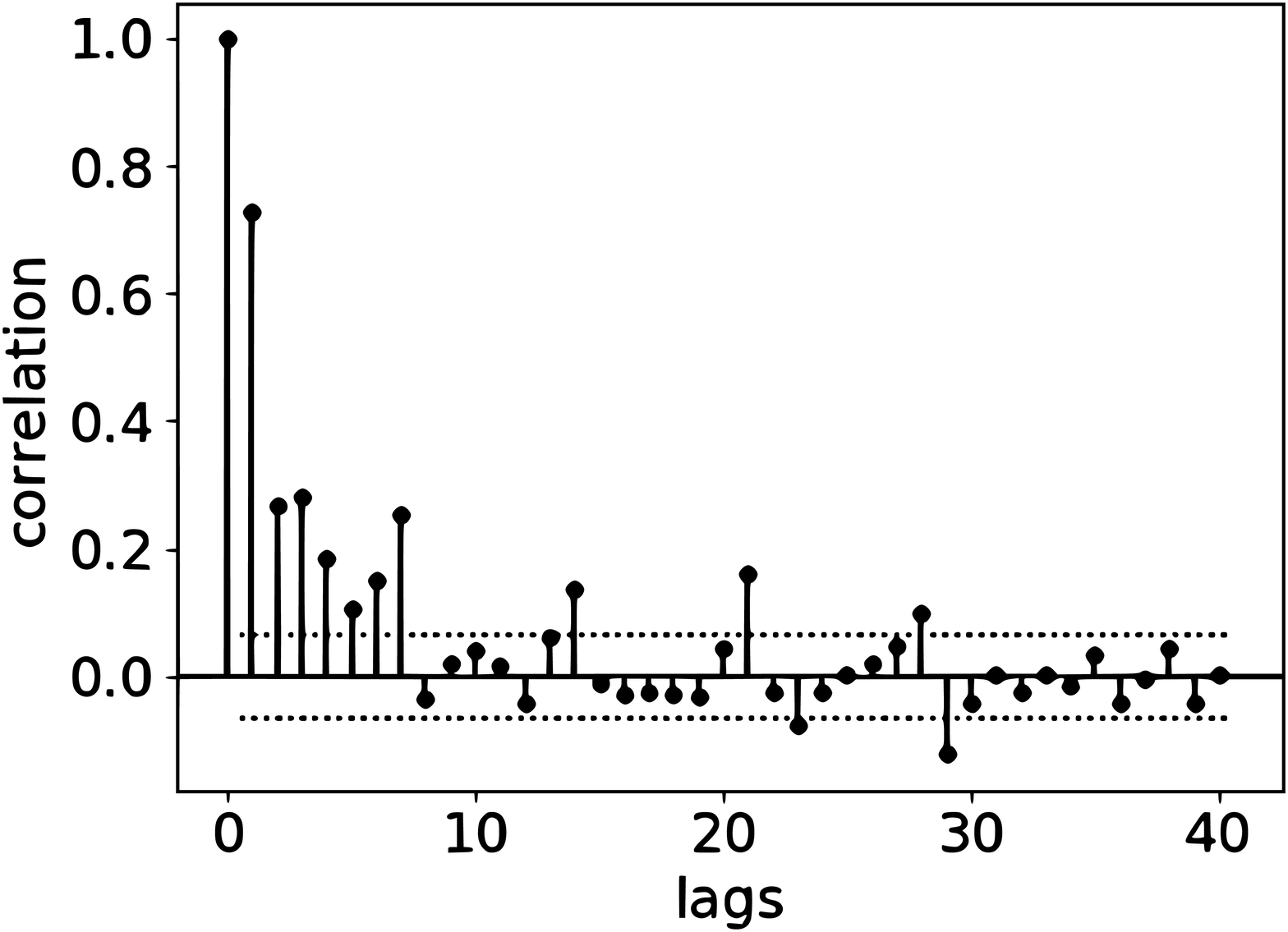}
\caption{Daily PU churning rate original series (top), its regular difference (middle), ACF (bottom left) and PACF (bottom right). Start of the training period is marked with a dashed line and corresponds to October 31, 2014. The dotted lines in the ACF and PACF define the significance region (for values larger than the positive line or smaller that the negative one) with 95\% confidence.}
\label{fig:ppd}
\end{figure}

\begin{figure}
\centering
\includegraphics[width=11cm]{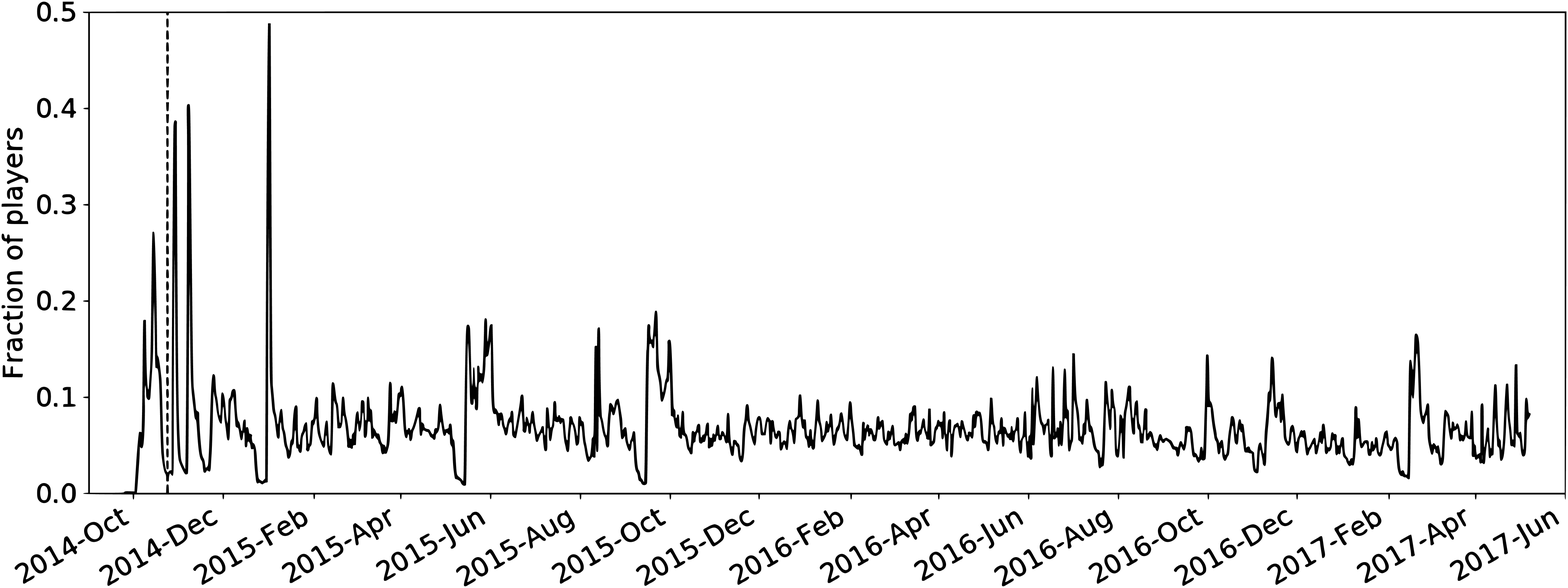}\\
\includegraphics[width=11cm]{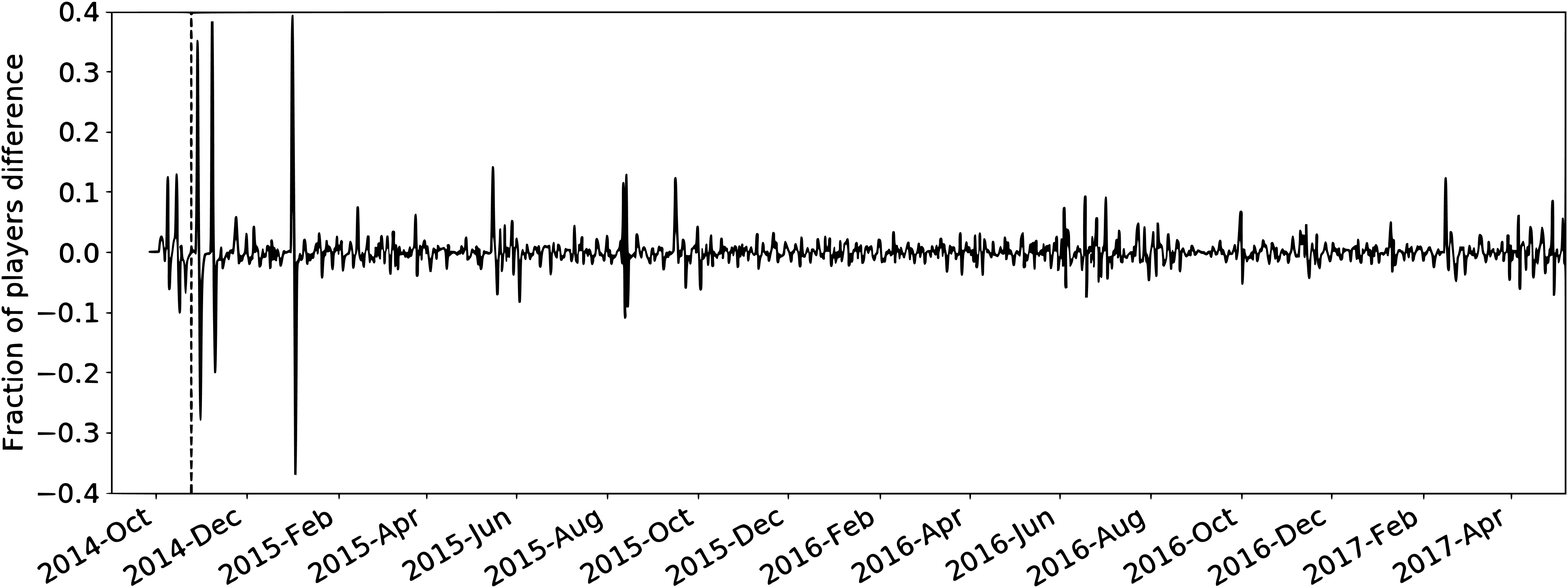}\\
\includegraphics[width=5.5cm]{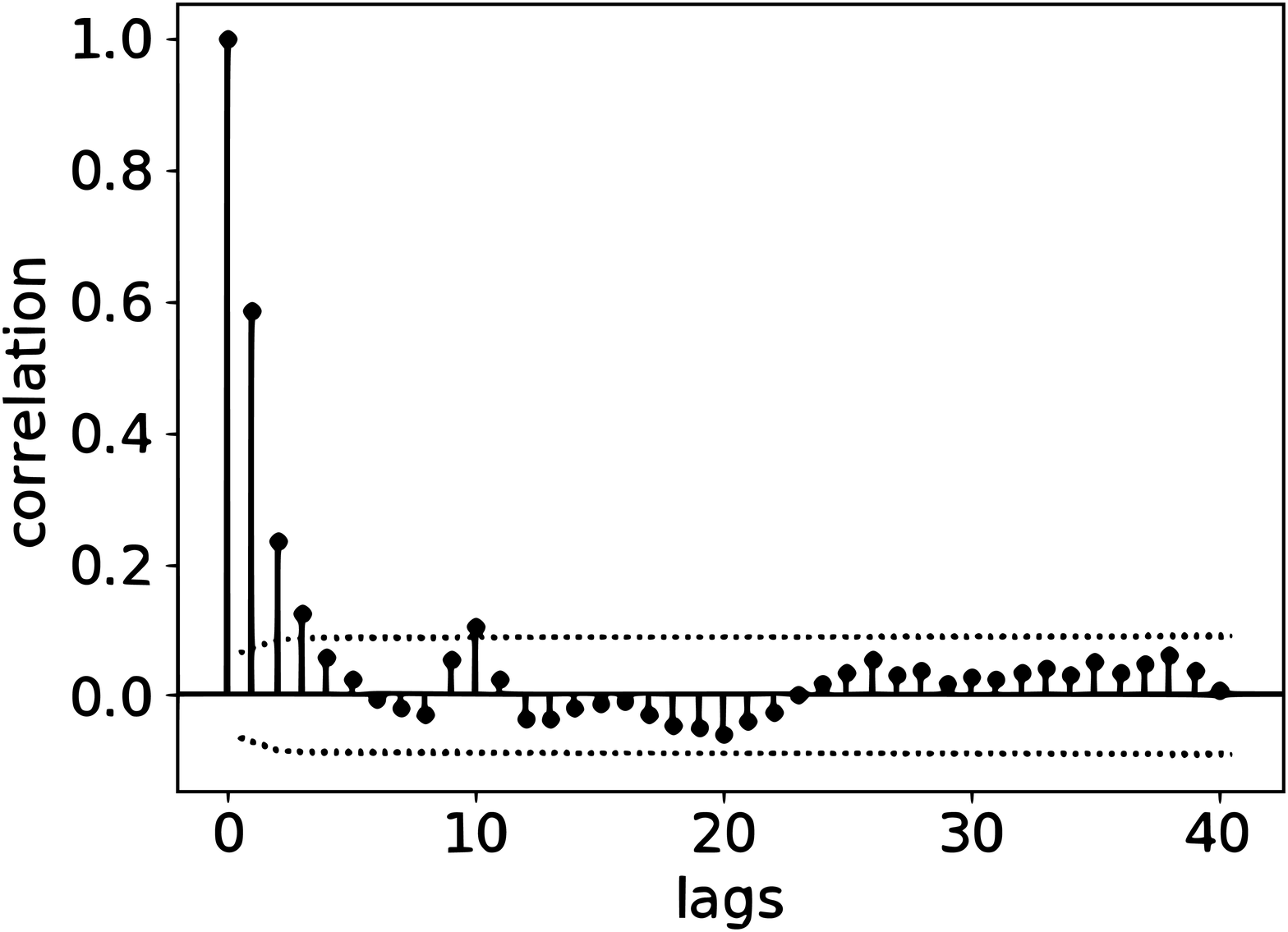}
\includegraphics[width=5.5cm]{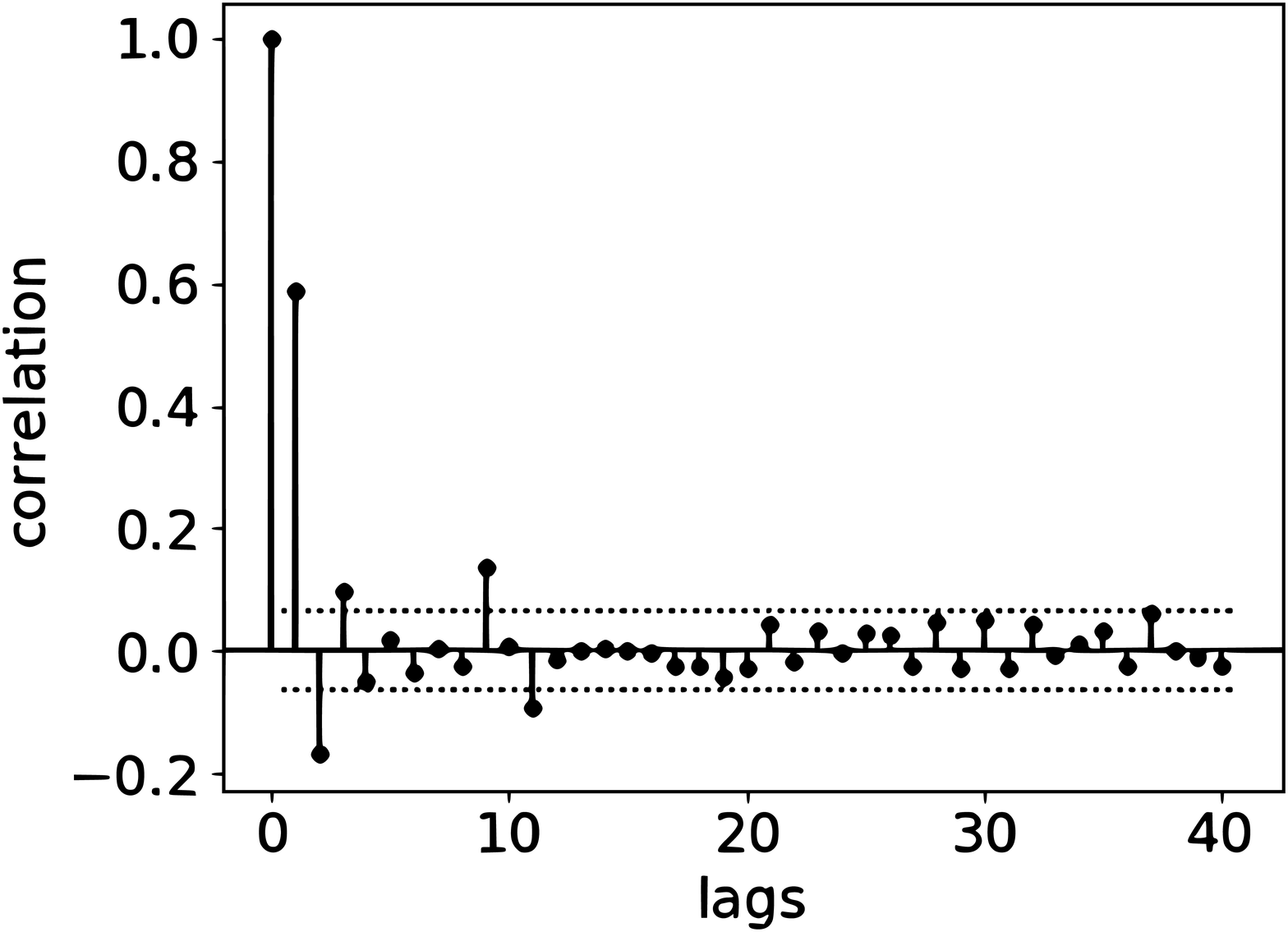}
\caption{Daily non-PU churning rate original series (top), its regular difference (middle), ACF (bottom left) and PACF (bottom right). Start of the training period is marked with a dashed line and corresponds to October 31, 2014. The dotted lines in the ACF and PACF define the significance region (for values larger than the positive line or smaller that the negative one) with 95\% confidence.}
\label{fig:pnd}
\end{figure}

\begin{figure}
\centering
\includegraphics[width=11cm]{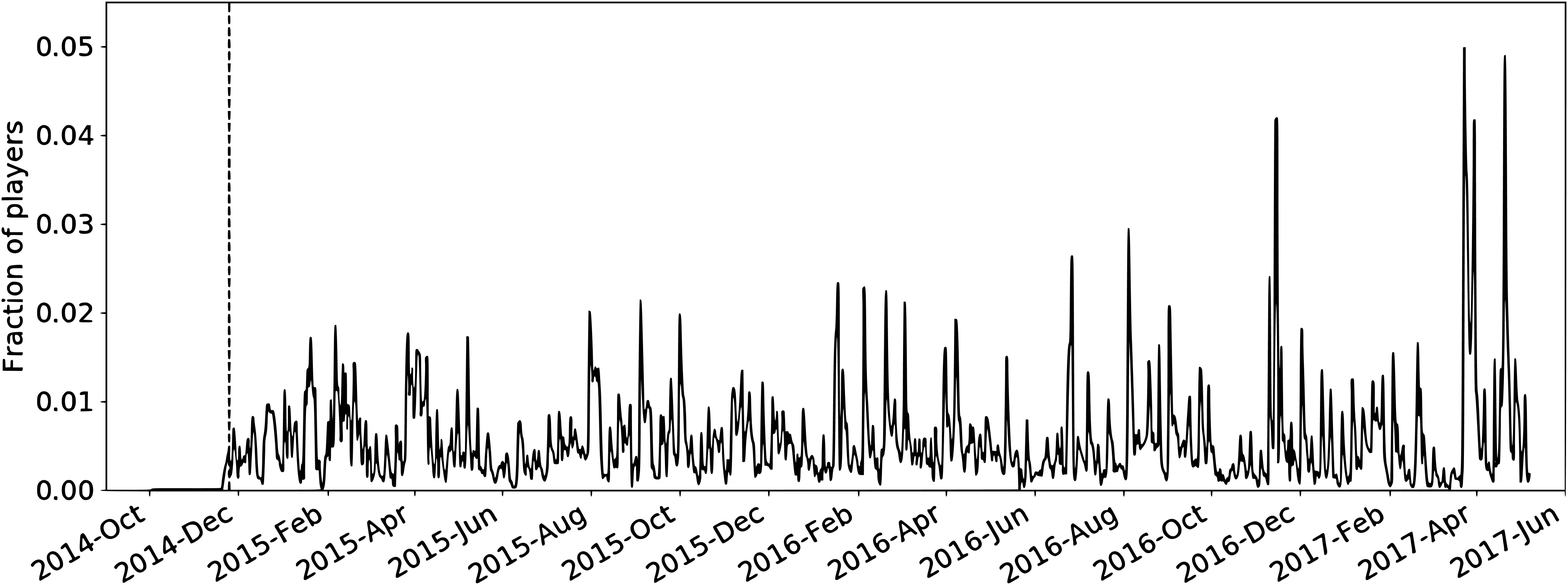}\\
\includegraphics[width=11cm]{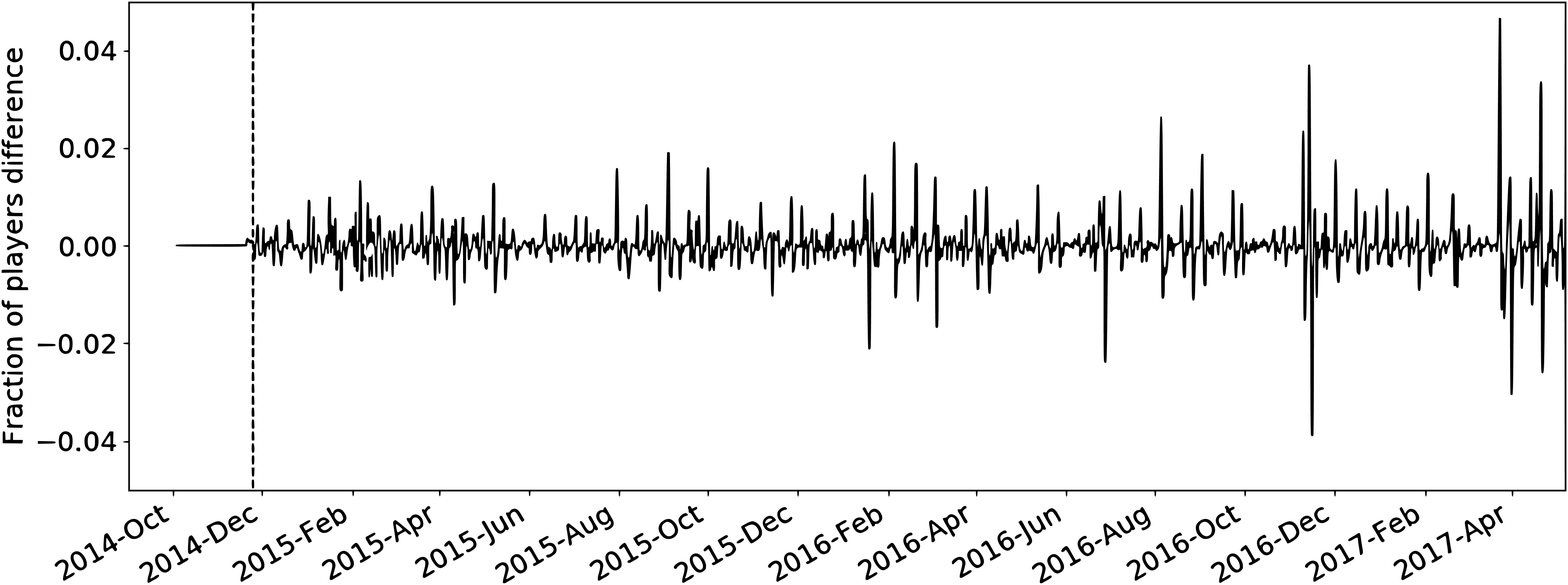}\\
\includegraphics[width=5.5cm]{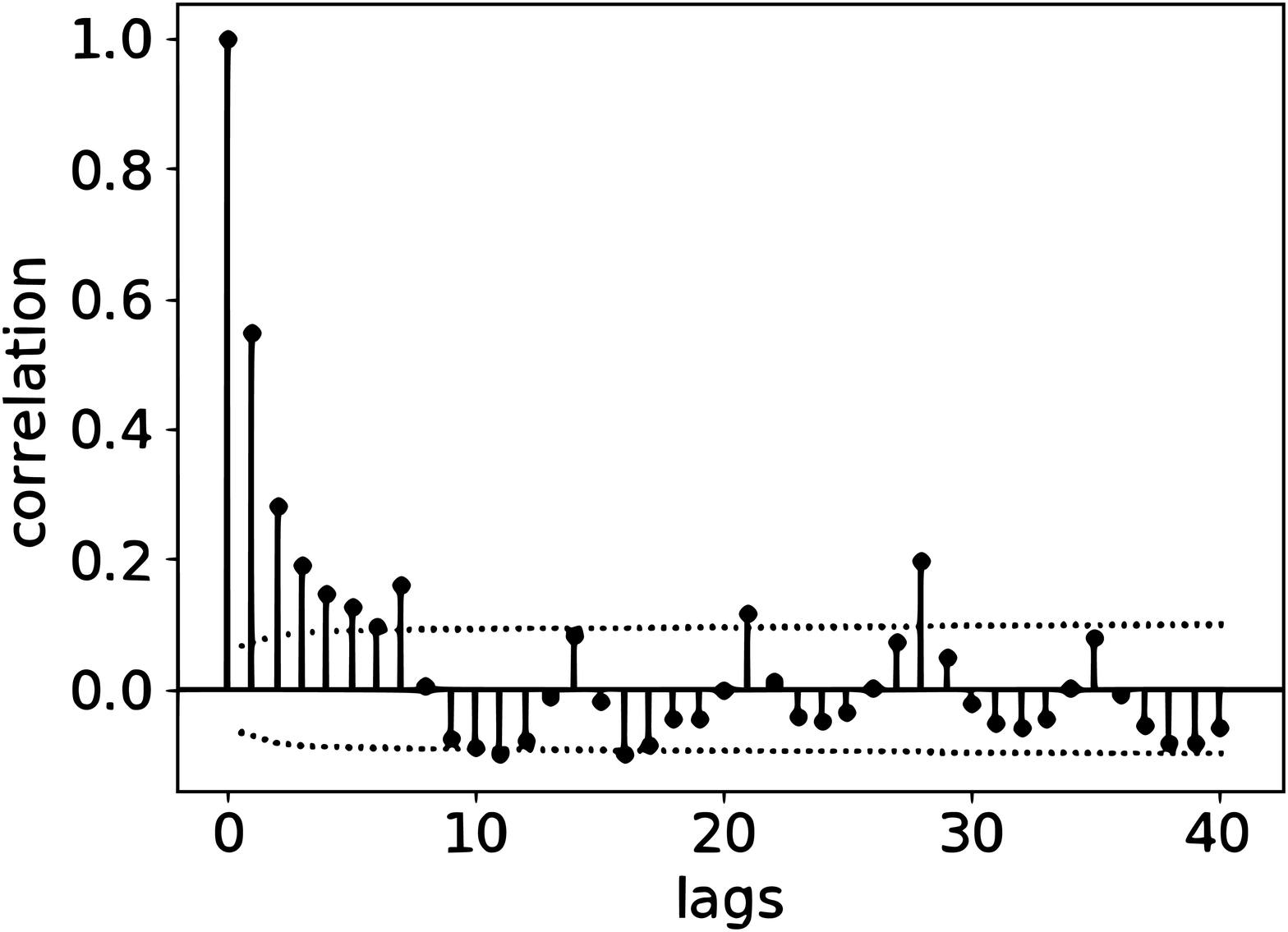}
\includegraphics[width=5.5cm]{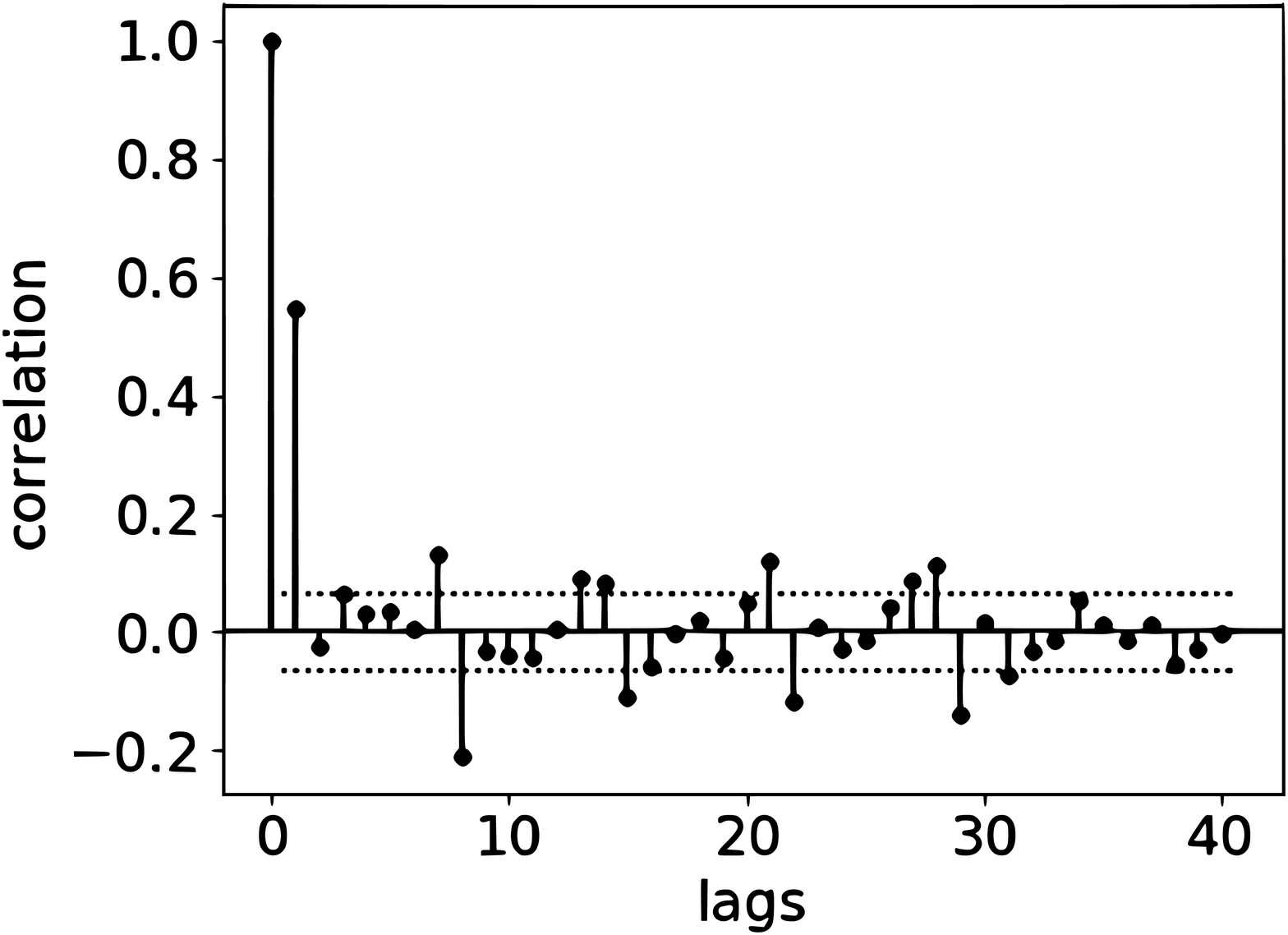}
\caption{Daily purchase churn rate original series (top), its regular difference (middle), ACF (bottom left) and PACF (bottom right). Start of the training period is marked with a dashed line and corresponds to November 5, 2014. The dotted lines in the ACF and PACF define the significance region (for values larger than the positive line or smaller that the negative one) with 95\% confidence.}
\label{fig:ppn}
\end{figure}

Figures \ref{fig:pnp}, \ref{fig:ppd}, \ref{fig:pnd} and \ref{fig:ppn} refer to conversion to PU, PU churn, non-PU churn and purchase churn rates respectively. They all show the original series at the top, the differenced series in the middle and the ACF and PACF of the original series (excluding the beginning of the series which is not used in the training and with regions outside the area delimited by the dotted lines corresponding to significant correlation values with 95\% confidence) at the bottom. Starting date for the training is shown as a dashed line on both the original and differenced series. The training began on October 5, 2014 and the ARIMA model used was (0, 1, 3) for conversion to PU. Churn modeling began on October 31, 2014 for both PU and non-PU, with (0, 1, 2) used as ARIMA for the former and (1, 1, 2) for the latter. For purchase churn the starting date was November 25, 2014 and the ARIMA chosen was (0,1, 3). As it was already mentioned, all UC models used a local level and weekly periodicity, and none (except new users) added a longer cycle component.

Not only are the same covariates selected for both models by following the process described in section \ref{sec:meth}, but the parameters estimated by both are very similar differing typically less than 10\% and in very few cases more than 20\%. The value estimated for a selection of parameters for the different series is displayed in Table \ref{tab:param-arima} for the ARIMA model and in Table \ref{tab:param-ll} for its UC counterparts. It is by no means comprehensive and it is meant to illustrate some of the discussions that follow only. 

Taking into account the multiplicative nature of the model used for the new users series (in that it is the log-transform of an absolute number), estimated parameters for it can be understood as elasticities. This means, for example, that using the ARIMA estimation, a day which is national holiday would mean nearly 5\% more new users than a day which is not.  For the rest of series that are conversion rates subject to an additive model, parameters should be understood as absolute increases. For example, this would mean that national holidays will make the probability of PU churn increase by 0.0033.

\begin{table*}[htb]
\caption{ARIMA estimates for a selection of parameters for the different series.\label{tab:param-arima}}
    \begin{tabular}{l c c c c c}\hline
    Parameter & New users & Conversion to PU & non-PU churn & PU churn & Purchase churn\\
    \hline
    National holidays & \num{4.89e-2} & \num{1.61e-4} & \num{3.30e-3} & - & \num{7.67e-4}\\
    Battle event (start) & - & \num{2.16e-4} & - & - & \num{-2.77e-4}\\
    Gacha 4 & - & \num{1.03e-3} & - & - & \num{1.61e-2}\\
    Raid event (start) & - & \num{-1.07e-3} & - & \num{3.80e-3} & \num{3.23e-3}\\
    
    Unknown 2017/02/09 & -  & \num{-1.33e-3} & \num{1.51e-2} &  \num{1.37e-4} & \num{1.88e-2}\\
    Marketing 2015/02/05-07 & \num{5.71e-3} & \num{9.97e-4} & \num{3.45e-2} & \num{4.38e-3}& \num{1.23e-3}\\
    Marketing 2015/03/16-25 & \num{6.06e-3} & - & \num{3.70e-2} & \num{5.14e-3}&-\\    
    Marketing 2015/05/25-31 & \num{6.07e-3} & - & \num{2.74e-3} & -&-\\
    Marketing 2016/09/21-22 & \num{6.25e-3} & \num{1.90e-4} & \num{2.40e-2} & - &-\\  
    Marketing 2017/02/07-09 & \num{2.06e-3} & - & \num{-1.26e-3} & - &-\\ 
    Promotion 2015/03/19 & - & \num{1.53e-3} & - & - & \num{4.60e-3} \\ 
    Promotion 2015/04/23-24 & - & \num{1.74e-3} & - & - & \num{2.30e-3} \\
    Promotion 2016/09/21-23 & - & \num{3.38e-3} & - & - & -\\
    \hline
    \end{tabular}
\end{table*}

\begin{table*}[htb]
 \caption{Local level estimates for a selection of parameters for the different series.\label{tab:param-ll}}
    \begin{tabular}{l c c c c c}\hline
    Parameter & New users & Conversion to PU & non-PU churn & PU churn & Purchase churn\\
    \hline
    National holidays & \num{5.17e-2} & \num{1.47e-4} & \num{2.83e-3} & - & \num{7.10e-4}\\
    Battle event (start) & - & \num{2.13e-4} & - & - & \num{-2.96e-4}\\
    Gacha 4 & - & \num{1.01e-3} & - & - & \num{1.59e-2}\\
    Raid event (start) & - & \num{-1.06e-3} & - & \num{3.88e-3} & \num{2.50e-3}\\
    
    Unknown 2017/02/09 & -  & \num{-1.25e-3} & \num{1.57e-2} &  \num{1.31e-4} & \num{1.79e-2}\\
     Marketing 2015/02/05-07 & \num{5.78e-3} & \num{9.81e-4} & \num{3.56e-2} & \num{4.17e-3}& \num{2.73e-3}\\
    Marketing 2015/03/16-25 & \num{6.45e-3} & - & \num{3.95e-2} & \num{5.17e-3}&-\\    
    Marketing 2015/05/25-31 & \num{5.49e-3} & - & \num{1.34e-3} & -&-\\
    Marketing 2016/09/21-22 & \num{6.12e-3} & \num{1.95e-4} & \num{2.29e-2} & - &-\\  
    Marketing 2017/02/07-09 & \num{2.72e-3} & - & \num{-1.88e-3} & - &-\\ 
    Promotion 2015/03/19 & - & \num{1.46e-3} & - & - & \num{3.44e-3} \\ 
    Promotion 2015/04/23-24 & - & \num{1.74e-3} & - & - & \num{2.11e-3} \\
    Promotion 2016/09/21-23 & - & \num{3.46e-3} & - & - & -\\
    \hline
    \end{tabular}
\end{table*}

The estimated weekly structure and of calendar and holiday effects is qualitatively similar for all conversion rates, and it has to do mainly with general patterns observed in the playtime. People tend to play more towards the end of the week and specially during weekends, and less on Monday to Wednesday. National holidays also have a clear positive impact in all series (except PU churn), while school holidays are not estimated as significant in any of the series, suggesting a limited amount of school age players in the game as compared to older working population.

In-game events have, as expected, no significant effect estimated in the new users series. None were also estimated significant in explaining non-PU churn, and only a couple of them had a low impact in PU churn. Most of them had however clear impact in conversion and purchase churn, suggesting they drive spending (or lack thereof) more than login engagement. Impact in both conversion to PU and purchase churn can be positive or negative, which implies that compared to no events at all, some encourage and some discourage spending. A typical event will have positive effect on both series (see for example \emph{Gacha 4} in tables \ref{tab:param-arima} and \ref{tab:param-ll}): they encourage spending and drive conversion to PU, but at least part of this effect is lost once the event is over, hence their positive delayed effect in purchase churn. Other event types have more interesting effects. \emph{Battle event} for example, not only drives conversion to PU, but actually reduces purchase churn. It does seem to motivate non-PU into becoming PU, while also driving expenditure in players who are already PUs. On the other hand, there are event types such as \emph{Raid event} that discourage conversion while having a positive effect in purchase churn and even a small effect in PU churn, which suggests that this event type discourages spending and that it is generally disliked by PUs.

Following the methodology described in section \ref{sec:meth}, using ARIMA models yields a reasonable marketing and promotion intervention scenario (as described in some more detail below). The same process using local level models however, as has already been briefly discussed, yielded worse results, with less interventions detected and producing degraded forecasts. Following the covariate selection process with the ARIMA defined interventions yields exactly the same selection for each series for both models. It was therefore decided to use only the ARIMA models for the intervention definition process. As already noted before, the poor performance of UC as compared to ARIMA models in outlier detection can be explained as the local level model's multiple noise terms are more able to absorb sudden changes in the series than the more fixed structure provided by ARIMA.

Marketing interventions have a large positive impact in the new users series, and typically also have a positive effect on both churn ones. Of course, campaigns with no effect (or even negative) in churn are precisely the most successful campaigns. Comparing the effect of the different new acquisition campaigns in these series gives an idea of which ones were targeting more effectively potential players that will do more than just try the game. A limited number of marketing interventions were also estimated as significant for the conversion to PU series. Either they were linked with something (for example new content) that encouraged spending, or they were simply good at motivating spending in people who were already players. Marketing type interventions were also tried with one and two days delay in the conversion to PU series. The idea was to account for newly acquired players through these campaigns that could decide to purchase for the first time in the next days of play. No significant impact was estimated for any of the marketing interventions, suggesting that this is not a frequent event, and that newly acquired players through these campaigns will either become PUs the same day they first log in, or will do so later on in the game if at all.

For example, most of the marketing interventions in Tables \ref{tab:param-arima} and \ref{tab:param-ll} have been chosen to have a comparable impact in the new users series (around a 6\% increase except for that taking place on February 2017). The four of them also impact non-PU churn and two of them impact PU churn as well. From these four, the most successful one would probably be that of May 2015: it has a much lower positive effect in non-PU churn and no effect in PU churn. The worst one would be that of March 2015: it has the highest impact on non-PU churn and also increases PU churn. Two of them also impact conversion to PU, and those on February 2015 even impact purchase churn. The final marketing intervention included (February 2017) is a perfect example of a very good campaign: it actually decreases non-PU churn, so not only has it attracted new users, it also seems to keep all (new and old) players engaged.

No promotion interventions or in-game covariates were found to be significant for the new users series as expected (only players would notice any of them). Many of the promotion interventions have also a noticeable impact in purchase churn (50 days later), but a lot of them do not, pointing at the promotions that made lasting conversions. For example, considering promotion interventions shown in Tables \ref{tab:param-arima} and \ref{tab:param-ll}, the most successful one would have been that of September 2016 (highest impact in conversion and no effect in purchase churn), while the least successful one that of March 2015. 

Only a handful of unknown interventions needed to be defined, most of them concerning purchase churn, which appears to be the series with dynamics less explained by the information available. Some of the interventions though, do seem to point at some clear effect negatively affecting engagement: they impact negatively conversion to PU while increasing both churn series and purchase churn. These could correspond to buggy releases or server failures that annoyed players. It could be the case of the unknown intervention shown in Tables \ref{tab:param-arima} and \ref{tab:param-ll} for February 9, 2017 that increases churn and purchase churn while negatively affecting PU conversion.

In regards to the actual marketing and promotion campaign planning revealed by the intervention definition process, it suggests that a lot was going on on both sides during the first months of the game, which makes sense after a new launch. After that, important new player acquisition campaigns seem to have run approximately every second month, except for the summer months that have campaigns on and off during the whole period. Promotion campaigns after the beginning of 2015 were shorter and sparser. This changed towards the end of 2016. Specially during 2017 there is an unprecedented length, intensity and frequency of promotion campaigns, with shorter and longer promotions constantly starting or ending. This also makes sense in that it seems to be an effort to maintain the total revenue generated by the game after it starts \emph{losing gas} (in-game sales, DAUs and total playtime start decreasing the second half of 2016, and these campaigns appear to manage to bring purchases back to their previous level).

\begin{figure}
\centering
\includegraphics[width=9cm]{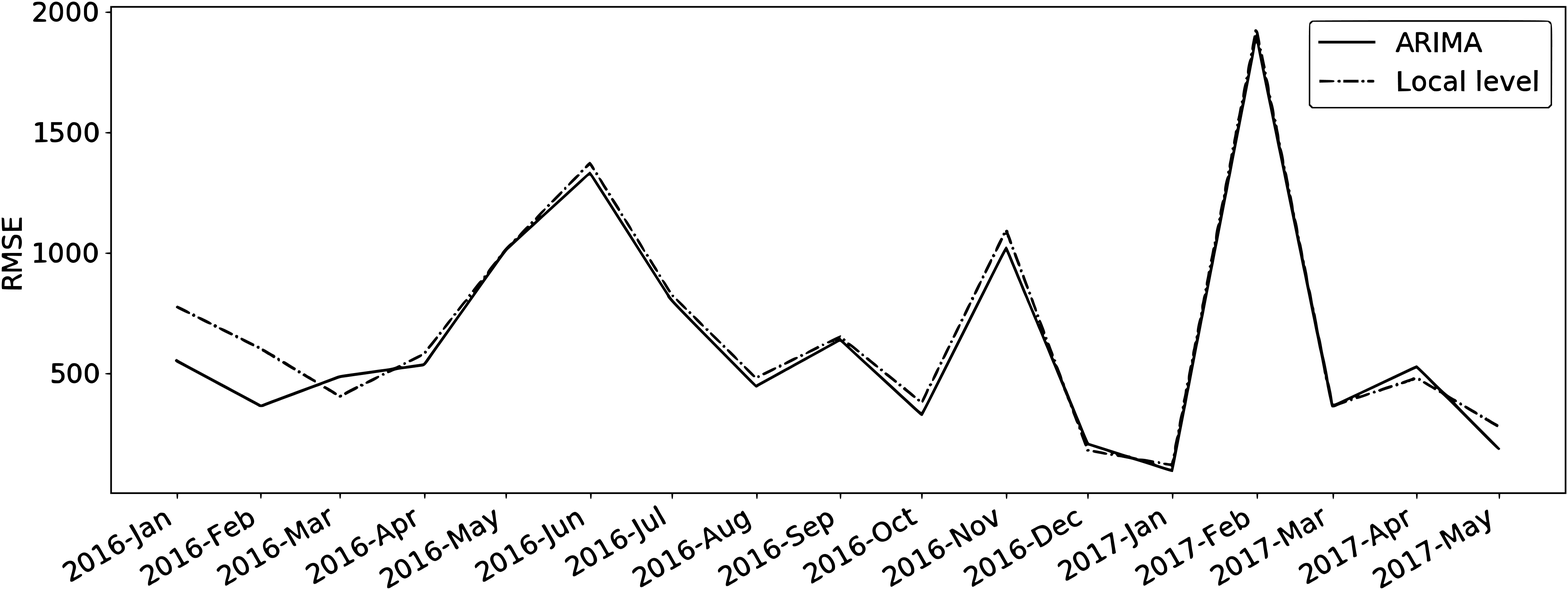}\\
\includegraphics[width=9cm]{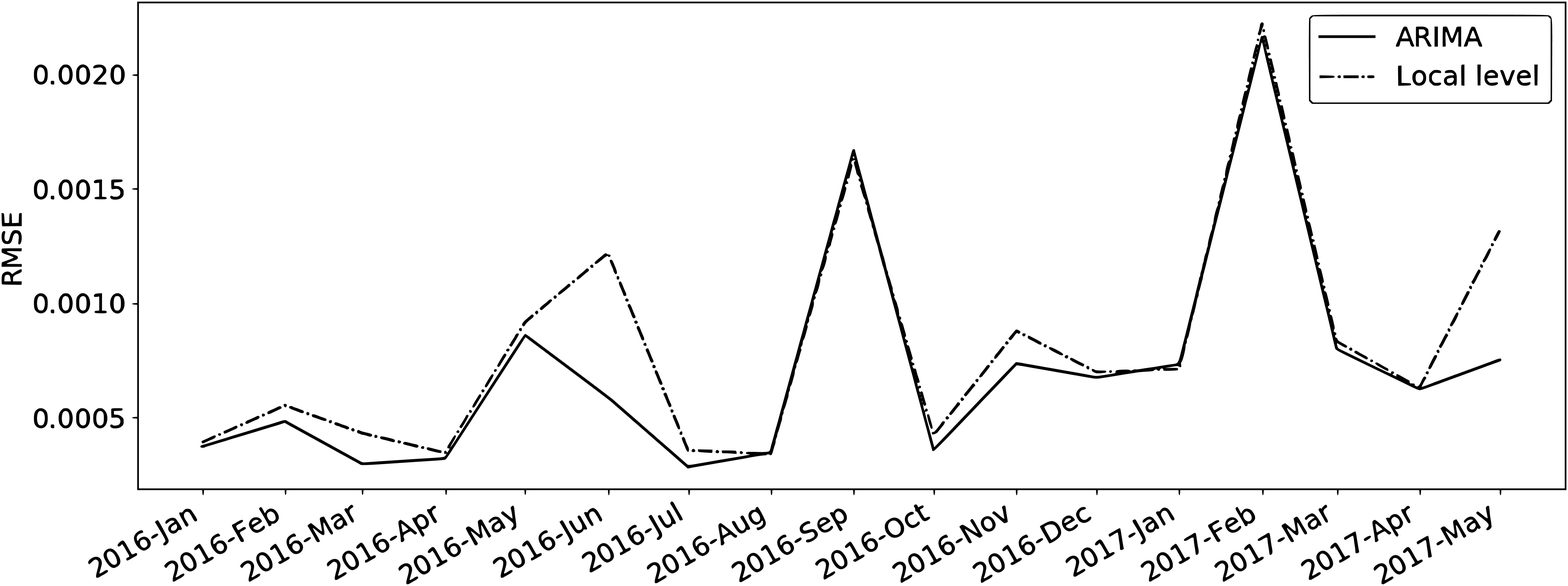}\\
\includegraphics[width=9cm]{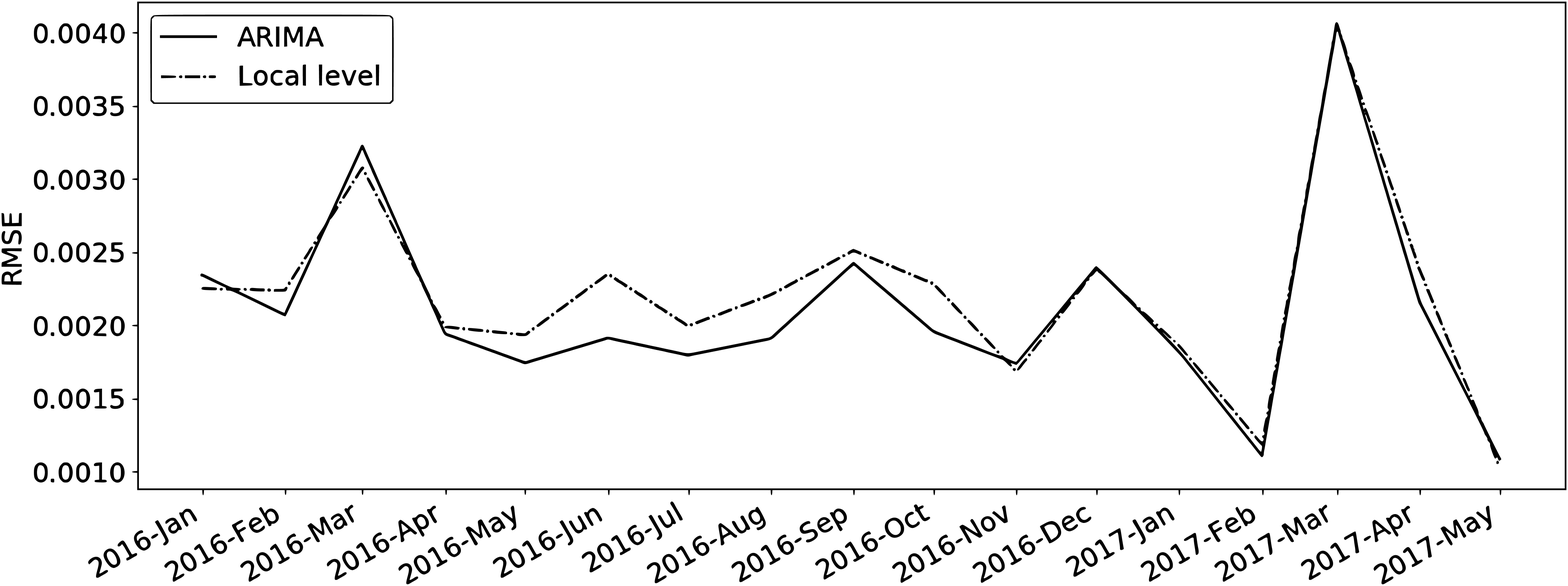}\\
\includegraphics[width=9cm]{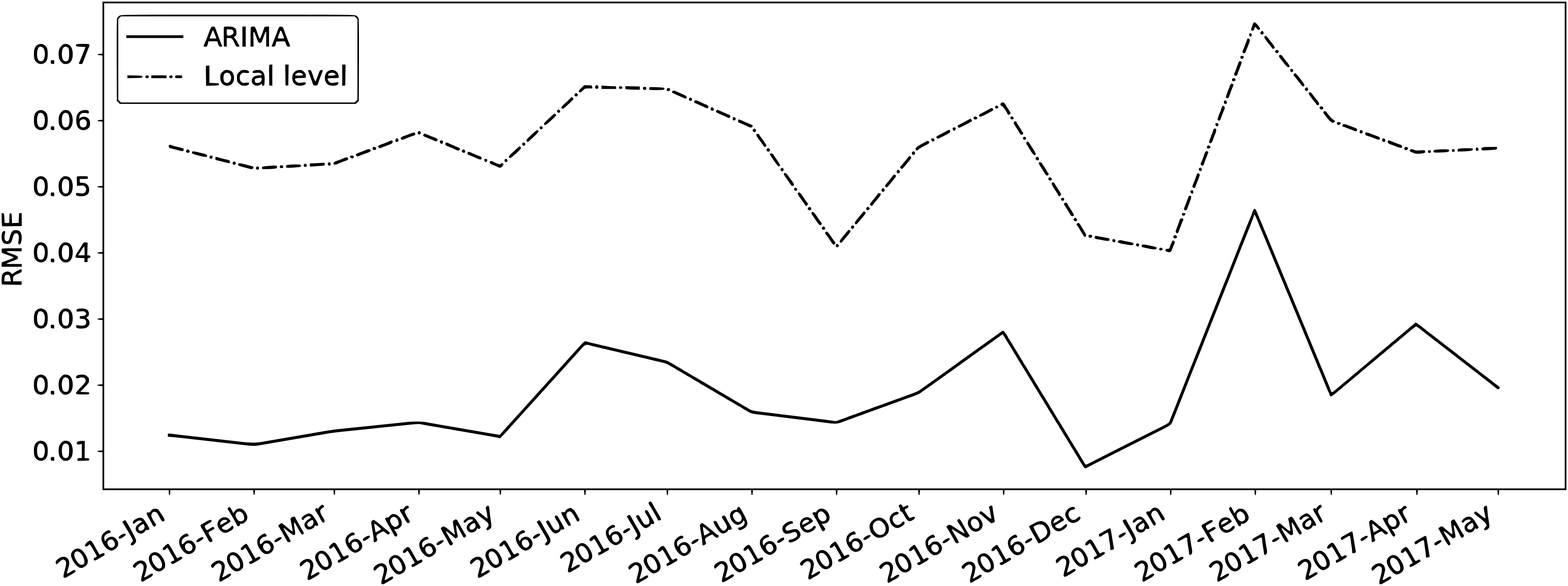}\\
\includegraphics[width=9cm]{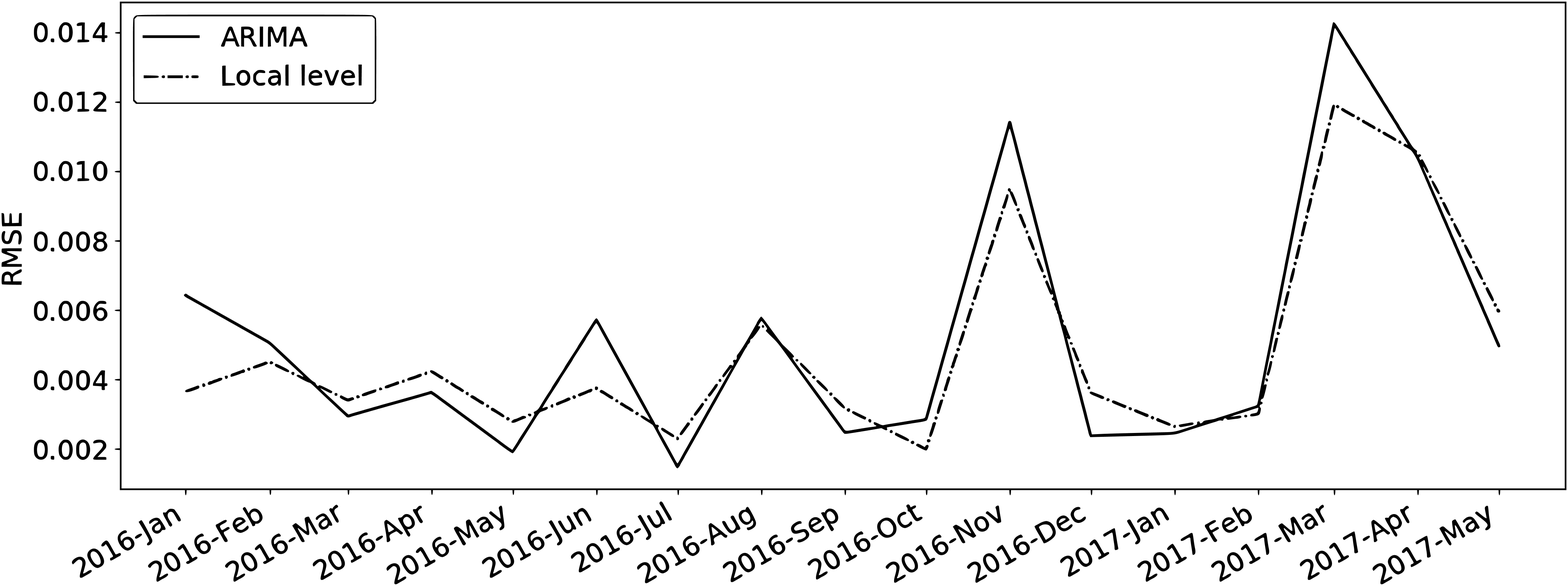}
\caption{Mean RMSE for all successive monthly forecasts for new users (top), conversion to PU (second row), PU churn (third row), non-PU churn (fourth row) and purchase churn (bottom). ARIMA is shown with a solid line and local level with a dashed lined.}
\label{fig:rmse}
\end{figure}

The mean values and standard deviation of the monthly MAE and RMSE of forecasts can be found in Tables \ref{tab:mae} and \ref{tab:rmse} respectively. Figure \ref{fig:rmse} shows the monthly values of RMSE for the daily forecasts produced by both models for each series. Note the goal of this validation metrics is merely to compare both models and to assess how their performance varies depending on the month and on the amount and importance of the missing interventions in the modeling.

\begin{table*}[hbt]
\caption{Monthly forecast MAE: mean and standard deviation (SD) for the ARIMA and local level models.\label{tab:mae}}
    \begin{tabular}{l c c c c}\hline
    Time Series & ARIMA Mean & ARIMA SD & Local Level Mean & Local Level SD\\
    \hline
    New users SD & 440.72 & 270.73 & 483.20 & 282.71\\
    Conversion to PU & \num{4.6e-4} & \num{2.5e-4} & \num{5.1e-4} & \num{2.5e-4}\\
    PU churn & \num{1.7e-3} & \num{5.8e-4} & \num{1.8e-3} & \num{5.8e-4}\\
    Non-PU churn & \num{1.4e-2} & \num{7.0e-3} & \num{5.2e-2} & \num{6.8e-3}\\
    Purchase churn & \num{3.3e-3} & \num{1.9e-3} & \num{3.3e-3} & \num{1.6e-3}\\
    \hline
    \end{tabular}
\end{table*}

\begin{table*}[hbt]
\caption{Mean RMSE for all successive monthly forecasts for new users (top), conversion to PU (second row), PU churn (third row), non-PU churn (fourth row) and purchase churn (bottom). ARIMA is shown with a solid line and local level with a dashed lined.\label{tab:rmse}}
    \begin{tabular}{l c c c c}\hline
    Time Series & ARIMA Mean & ARIMA SD & Local Level Mean & Local Level SD\\
    \hline
    New users SD & 634.64 & 461.44 & 677.91 & 463.86\\
    Conversion to PU & \num{7.1e-4} & \num{5.0e-4} & \num{8.2e-4} & \num{5.2e-4}\\
    PU churn & \num{2.1e-3} & \num{7.0e-4} & \num{2.2e-3} & \num{6.8e-4}\\
    Non-PU churn & \num{1.9e-2} & \num{9.4e-3} & \num{5.6e-2} & \num{8.9e-3}\\
    Purchase churn & \num{5.1e-3} & \num{3.7e-3} & \num{4.9e-3} & \num{3.0e-3}\\
    \hline
    \end{tabular}
\end{table*}

As Figure \ref{fig:rmse} shows, forecast accuracy is comparable but tends to be slightly worse for the local level models as compared to ARIMA, except in the case of non-PU churn where its performance is significantly worse. This could be related to its much higher local level signal-noise ratio (ratio between the local level and the irregular or residual variances) of this model as compared to that of other series. Higher values mean less lags into the past observations are taken into account for the forecasting \cite{harvey2006}, which could make the correct level of the series harder to capture. High peaks in Figure \ref{fig:rmse} correspond to months with many days of and/or high impacting campaigns, whose effect will be captured by interventions once that month becomes part of the training period.  

In general these findings match those found by Harvey \cite{harvey2006} and Andrews \cite{andrews} (which offer discussions on the similarities and differences of ARIMA and UC, and a comparison of their performance). The results are typically very similar, while UC models are simpler and need much less human intervention. The difference for non-PU churn might arise from the daily nature of the series (as compared to the weekly and monthly series examined in the references noted above). This could account for a larger number of outliers (that are not smoothed out through average or aggregation), that could in turn degrade the forecasts significantly particularly for large signal-noise ratios. Studying how this type of modeling behaves depending on the time granularity used seems like an interesting path to explore.

%%%%%%%%%%%%%%%%%%%%%%%%%%%%%%%%%%%%%%%%%%%%%%%%%%%%%%%%%%%%%%%%%%%%%%%%%%%%%%%%%%%%%%%%%%
\section{Summary and conclusions}
\label{sec:conc}

Time series modeling of conversion rates between different groups of players can provide insights into the effect of different in-game and external events, and a way of detecting relevant missing information. It is particularly a very interesting approach to understand how different elements of in-game dynamics affect differently different types of players.

Two different SSM approaches were considered and compared: ARIMA and UC. The latter is more robust and needs much less human intervention in regards to model definition, while still providing the same covariate selection and very similar parameter estimation than the statistically more complex ARIMA counterpart. If the only or main interest is to classify in-game or external events in terms of their effects on the different conversion rates, in order to better understand game dynamics and improve game planning, a local level UC model would be the preferred option. If the focus is predictive accuracy, the picture varies slightly. While both models typically have very similar forecasting performance, ARIMA tends to do better in more cases, and can very drastically outperform local level models for some transitions (non-PU churn in this case). Finally, if the detection of missing campaigns or other pieces of missing relevant information is needed, again here the ARIMA models do a much better job, providing a plausible scenario consistent across all transition rates.

Both models (and particularly so UC) are not costly to re-estimate, so a plausible setup would be to combine the monthly or weekly forecasts for planning and resource allocation, with daily updated estimations when a new data point is available (and possibly updated forecasts if these could be useful for updated planning). This would allow for early detection of anything unusual going on (e.g. server failures or buggy releases). This anomalous behaviour detection could be carried out using the residuals at the end of the estimated series (large absolute values or several successive values of the same signed) and/or large discrepancies between very short range forecasts and actual observed values. This process could be fully automatic, raising alarms only when unexpected deviations occur. It would also provide an immediate assessment of how well or bad are in-game events doing as compared to similar ones in the past.

Future extensions of this work will include increasingly complex player type landscapes, trying additional time series models to evaluate their performance, and assessing the impact of using the resulting forecasts in individual player behavior modeling.  

%%%%%%%%%%%%%%%%%%%%%%%%%%%%%%%%%%%%%%%%%%%%%%%%%%%%%%%%%%%%%%%%%%%%%%%%%%%%%%%%%%%%%%%%%%
\section{Software used}
\label{sec:software}
All analysis and predictions were performed with Python 3.6.7, making use of the datetime, numpy \cite{numpy}, pandas \cite{mckinney}
and statsmodels \cite{seabold2010statsmodels} libraries. Plots were produced using the matplotlib \cite{Hunter:2007} library.

\nocite{*} 
\bibliographystyle{ios1}           % Style BST file.
\bibliography{main}        % Bibliography file (usually '*.bib')

\end{document}